\documentclass[12pt]{article}
\usepackage{amsmath}
\usepackage{amsfonts}
\usepackage{amssymb}
\usepackage{float}
\usepackage{ulem}
\usepackage{graphicx}
\usepackage{slashed}

\usepackage{color}
\usepackage{longtable}
\restylefloat{table}

\newlength{\abstractwidth}

\usepackage{epsf}
\usepackage{cancel}

\pdfoutput=0

\renewcommand{\thanks}[1]{\footnote{#1}}

\renewcommand{\theequation}{\thesection.\arabic{equation}}
\newcommand{\bea}{\begin{eqnarray}}
\newcommand{\eea}{\end{eqnarray}}
\newcommand{\ee}{\end{equation}}
\newcommand{\be}{\begin{equation}}

\newcommand{\ea}{\end{array}}
\newcommand{\bac}{\begin{array}{c}}
\newcommand{\bacc}{\begin{array}{cc}}
\newcommand{\barcl}{\begin{array}{r@{}c@{}l}}
\newcommand{\brcl}{\begin{array}{rcl}}
\newcommand{\bdm}{\begin{displaymath}}
\newcommand{\edm}{\end{displaymath}}

\def\cN{{\cal N}}

\def\a{\alpha}
\def\b{\beta}

\def\no{\nonumber}

\def\nvec5{{\tilde n}}

\def\Tr{{\rm Tr}}
\def\id{\protect{{1 \kern-.28em {\rm l}}}}

\def\spa#1.#2{\left\langle#1\,#2\right\rangle}
\def\spb#1.#2{\left[#1\,#2\right]}

\usepackage{color}

\renewcommand{\theequation}{2.\arabic{equation}}
\begin{document}
 
\textwidth 170mm
\textheight 230mm
\topmargin -1cm
\oddsidemargin-1cm \evensidemargin -1cm
\topskip 9mm
\headsep9pt

\overfullrule=0pt
\parskip=2pt
\parindent=12pt
\headheight=0in \headsep=0in \topmargin=0in \oddsidemargin=0in

\vspace{ -3cm} \thispagestyle{empty} \vspace{-1cm}

\begin{flushright}
UUITP-57/18 \hskip 1cm
\end{flushright}


\begin{center}

{\Large \bf One-loop Amplitudes for $\cN=2$ Homogeneous \\[4pt] Supergravities}\\

\bigskip

\bigskip

{\large 
	Maor Ben-Shahar and Marco Chiodaroli}

\bigskip

\bigskip

 Department of Physics and Astronomy \\ Uppsala University \\
	SE-75108 Uppsala, Sweden\\
	\texttt{ marco.chiodaroli,benshahar.maor@physics.uu.se}

\bigskip

\bigskip

\bigskip

\end{center}

\begin{abstract}
We compute one-loop matter amplitudes in homogeneous Maxwell-Einstein supergravities with $\cN=2$ supersymmetry using the double-copy construction.
We start from  amplitudes of $\cN=2$ super-Yang-Mills theory with matter that obey manifestly the duality between color and kinematics. Taking advantage of the fact that amplitudes with external hypermultiplets have kinematical numerators which do not present any explicit dependence on the loop momentum, we  find a relation between the one-loop divergence of the supergravity amplitudes and the beta function of the non-supersymmetric gauge theory entering the  construction.    
Two distinct linearized counterterms are generated at one loop. The divergence corresponding to the first is nonzero for all homogeneous supergravities, while the divergence associated to the second vanishes only in the case of the four Magical supergravities. 
\end{abstract}

\baselineskip=16pt
\setcounter{equation}{0}
\setcounter{footnote}{0}

\newpage

\tableofcontents

\section{Introduction and summary of results}
\renewcommand{\theequation}{1.\arabic{equation}}
\setcounter{equation}{0}

In the past decade, the duality between color and kinematics and the related double-copy construction \cite{Bern:2008qj,Bern:2010ue}  have produced a breath-taking increase in our ability to conduct multi-loop calculations in gravity and supergravity theories. 
The double-copy method expresses loop-level gravity integrands in terms of gauge-theory building blocks, which are typically far easier to obtain.
Taking advantage of this construction, 
calculations have been performed up to five loops in maximal supergravity \cite{Bern:2017yxu,Bern:2017ucb,Bern:2018jmv} and up to four loops in half-maximal supergravity \cite{Bern:2012cd,Bern:2012gh,Bern:2013uka}. 
The studies of ultraviolet (UV) behaviors which have been made possible by these calculations have
uncovered a set of so-called enhanced cancellations \cite{Bern:2014sna,Bern:2017lpv} and  
 revealed a connection between UV-divergences and $U(1)$ quantum anomalies \cite{Carrasco:2013ypa,Bern:2017rjw}.

Going beyond very special theories with extended supersymmetry, it has become clear that there exists a large web of double-copy-constructible theories, which includes pure ungauged supergravities \cite{Bern:2011rj,Johansson:2014zca}, ungauged supergravities with various matter contents \cite{Carrasco:2012ca,Bern:2013yya,Chiodaroli:2015wal,Anastasiou:2016csv,Anastasiou:2017nsz,Johansson:2017bfl}, Yang-Mills-Einstein theories \cite{Chiodaroli:2014xia,Chiodaroli:2016jqw}, spontaneously-broken supergravities \cite{Chiodaroli:2015rdg}, gauged supergravities \cite{Chiodaroli:2017ehv} and conformal supergravities \cite{Johansson:2017srf,Johansson:2018ues}. 
Conversely, color/kinematics duality has been established for large classes of gauge theories, including for example QCD \cite{Johansson:2015oia, delaCruz:2015dpa}.
Various non-gravitational theories, most prominently the Dirac-Born-Infeld theory and the non-linear sigma model, can also be understood in terms of the double copy and can provide examples in which the duality between color and kinematics is particularly well-understood \cite{Chen:2013fya, Chen:2014dfa, Cheung:2016prv, Du:2016tbc, Chen:2016zwe}.    
In addition, the double copy has been instrumental in explaining amplitude relations between Einstein-Yang-Mills and Yang-Mills (YM) theories \cite{Chiodaroli:2017ngp} (see also \cite{Stieberger:2016lng, Stieberger:2014cea, Stieberger:2015qja, Nandan:2016pya, delaCruz:2016gnm, Teng:2017tbo, Du:2017gnh,Nandan:2018ody}).

In the case of gravitational theories that can be directly obtained from  string theory, insight about color/kinematics duality can be obtained from string monodromy relations \cite{Stieberger:2009hq,BjerrumBohr:2009rd,BjerrumBohr:2010zs,Tourkine:2016bak,Hohenegger:2017kqy,Fu:2018hpu}.
Duality-satisfying structures can be obtained from the low-energy limit of string theory amplitudes \cite{Mafra:2011kj,Mafra:2014gja, He:2015wgf, Mafra:2015mja}. 
New sets of double-copy relations that combine string and field-theory building blocks have also become a powerful tool for understanding the structure of various string theory amplitudes \cite{Stieberger:2013wea,Mafra:2017ioj,Huang:2016tag,Azevedo:2018dgo}.  
In addition, the double copy is an intrinsic feature of various modern approaches to scattering amplitudes, including the CHY formalism \cite{Cachazo:2013hca, Cachazo:2013iea, Cachazo:2013gna, Cachazo:2014nsa, Cachazo:2014xea,Bjerrum-Bohr:2016axv,Fu:2018hpu} and ambitwistor strings \cite{Mason:2013sva,Casali:2015vta}.

Given the considerable progress in establishing double-copy structures for wider and wider classes of theories, it is natural to seek additional examples in which the double copy can be employed for shedding some light on the loop-level properties of supergravities with matter. A related issue is that the double-copy is by nature a construction at the integrand level, a fact that makes it more difficult to establish a direct relation between the physical properties of a gravity theory and the ones of the gauge theories entering the construction, which become manifest after integration.   

In this paper we focus on the double-copy construction for Maxwell-Einstein supergravities with $\cN=2$ supersymmetry in four and five dimensions and homogeneous target spaces, which was first presented in ref. \cite{Chiodaroli:2015wal}. These theories have been explicitly classified in the supergravity literature \cite{deWit:1991nm} and provide a natural testing ground for amplitude methods. Their double-copy construction involves a $\cN=2$ super-Yang-Mills (sYM) theory with one half-hypermultiplet in a pseudo-real representation as one gauge-theory factor, and a non-supersymmetric YM theory with adjoint scalars and matter fermions in the same representation as the other gauge-theory factor. 

Loop amplitudes with four external hypermultiplets that obey the duality between color and kinematics were first obtained in ref. \cite{Chiodaroli:2013upa}. These amplitudes have the striking property that the loop momentum does not appear in the kinematical numerators. Focusing on supergravity amplitudes between four vector multiplets in which the vectors are constructed as the product of two spin-$1/2$ asymptotic states, the above property allows us to find an explicit relation between supergravity  amplitudes and  gauge-theory physical quantities which holds at the integrated level. Our main result is that the one-loop divergence for the (super)amplitudes we are inspecting can be directly related to a linear combination of various parts of the one-loop beta function of the non-supersymmetric gauge theory entering the construction, 
\begin{eqnarray}
{\cal M}^{\text{1-loop}}\Big|_{\text{div}} \! \! \! \! &=& \! - {   s \delta^4(Q) \over \langle 12 \rangle \langle 34 \rangle   } \Big({\kappa \over 2}\Big)^4 
\left\{ s A^{\text{tree}}_{s,\phi} \Big( \beta_\phi \big|_{T(G)} \! - \! { \beta_\phi \over 2} \Big|_{T(R)} \Big) 
\! + \! s A^{\text{tree}}_{s,A}  \Big( \beta_A \big|_{T(G)} \! - \! { \beta_A \over 2} \Big|_{T(R)} \Big) 
\right\} {c_\Gamma \over \epsilon} \no \\ && \hskip 10cm + \text{Perms}, 
\end{eqnarray}
where $\beta_\phi, \beta_A$ are the beta functions for the gauge and Yukawa couplings and $\beta\big|_{T(G), T(R)}$ are the parts of the beta functions proportional to the index of the adjoint and  matter representations. $A^{\text{tree}}_{s,A}$ and $A^{\text{tree}}_{s,\phi}$ are the contribution to the non-supersymmetric gauge theory's tree-level $s$-channel from gluon and scalar exchange, respectively. The presence of the Grassmann delta function in the above formula makes $\cN=2$ supersymmetry manifest. 

This paper is structured as follows. In Section \ref{sec2}, we review the construction for homogeneous supergravities and introduce the gauge-theory Lagrangians employed in the calculation. In Section \ref{sec3}, we consider one-loop gauge-theory amplitudes with external hypermultiplets and extend the results of ref. \cite{Chiodaroli:2013upa} to the pseudo-real case by restoring the permutation symmetry at the superamplitude's level.
In Section \ref{sec4}, we compute one-loop amplitudes for the non-supersymmetric gauge theory and use the double-copy method for obtaining gravity amplitudes. We then conduct multiple checks and comparisons with previous results in the literature and end with a discussion of future directions. 

\section{Double-copy construction for homogeneous supergravities \label{sec2}}
\renewcommand{\theequation}{2.\arabic{equation}}
\setcounter{equation}{0}

The double-copy construction relies on organizing gauge theory amplitudes at $L$ loops and $n$ points  as sums over a set of cubic graphs, 
\begin{equation}
{\cal A}^{(L)}_{n} = i^{L-1} g^{n-2+2L} \sum_{i \in \text{cubic}}\, \int \frac{d^{LD}\ell}{(2\pi)^{LD}} \frac{1}{S_i} \frac{C_{i} n_i}{D_i} \, ,
\label{gaugepresentation}
\end{equation}
where $D_i$ are products of 
inverse scalar propagators and $S_i$ are symmetry factors. Each cubic graph is associated to a color factor
$C_i$, which is constructed out of gauge-group invariant tensors (i.e., structure constants, representation matrices and Clebsch-Gordan coefficients with three matter-representation indices). Since the relevant group-theoretical objects obey algebraic relations (e.g. commutation relations and  Jacobi identities), there exist  triplets of graphs
$\{i,j,k\}$ such that $C_i + C_j + C_k = 0$. 
Graphs in the above amplitude presentation are further associated to numerator factors, 
which contain theory-specific kinematical information and involve external and loop momenta, as well as Grassmann variables in case of superamplitudes. 
A theory obeys color/kinematics duality if the amplitude numerators possess the same algebraic properties as the corresponding color factors,
\begin{equation}
n_i - n_j = n_k  \quad \Leftrightarrow \quad  C_i - C_j=C_k \, .
\label{duality}
\end{equation}

\smallskip

The double-copy construction utilizes duality-satisfying numerators to write (super)gravity amplitudes by replacing the color factors with a second set of numerators, 
\be
\mathcal{M}^{(L)}_n \!= \!i^{L-1}\Big(\frac{\kappa}{2}\Big)^{n-2+2L} \!\!\!\! \sum_{i\in {\rm cubic}} \! \int \!
\frac{d^{LD}\ell}{(2\pi)^{LD}}\frac{1}{S_i}\frac{n_i\tilde{n}_i}{D_i},
\label{Sugra}
\ee
where $\kappa$ is the gravity coupling. The numerators $n_i,\tilde{n}_i$ may come from different gauge theories which share the same set of color factors. 
The formula requires that at least one of the two sets of numerators satisfies
manifestly color/kinematics duality \cite{Bern:2008qj,Bern:2010ue}. If this requirement is satisfied, it is possible to show that the double-copy amplitudes obey the Ward identities corresponding to linearized diffeomorphisms and hence have the interpretation of amplitudes from some gravitational theory \cite{Chiodaroli:2017ngp}. In general, identifying the gravity theory produced by a given double-copy construction (i.e. reconstructing the corresponding Lagrangian) is a highly-nontrivial problem. However, symmetry considerations, combined with minimal information on the supergravity interactions, have been instrumental in applying the double-copy method to very large classes of theories. 

Among the double copies formulated to date, the construction for $\cN=2$ homogeneous Maxwell-Einstein supergravities plays a key role. Supergravities in this family have been explicitly classified by the supergravity literature \cite{deWit:1991nm} and hence give a natural testing ground for the double-copy method. 
Their double-copy construction has the following schematic form \cite{Chiodaroli:2015wal}:
\begin{equation}
\Big( \cN=2  \text{ hom. sugras } \Big) = \left( \cN=2 \text{ sYM } + {1 \over 2} \text{ hyper}  \right) 
\otimes \left( \text{YM}_D + P \text{ fermions} \right) \ .\qquad \ 
\end{equation}
The right gauge theory entering the construction is a YM theory in $D$ spacetime dimensions with $P$ additional fermions which transform in a pseudo-real gauge group representation. The corresponding Lagrangian is 
\begin{equation}
{\cal L}_R = -{1 \over 4} F^{\hat a}_{\mu \nu} F^{\hat a \mu \nu} + {i \over 2} \bar \lambda \Gamma^\mu D_\mu  \lambda  \ ,  \label{L-left}
\end{equation}
where field strengths and covariant derivatives are defined as 
\begin{eqnarray}
F^{\hat a}_{\mu \nu} &=& \partial_\mu A^{\hat a}_\nu - \partial_\nu A^{\hat a}_\mu + g f^{\hat a \hat b \hat c } A^{\hat b}_\mu A^{\hat c}_\nu \ , \\
D_\mu \lambda &=& \partial_\mu \lambda  - i g T_R^{\hat a} A_\mu^{\hat a} \lambda \ .
\end{eqnarray}
Here $\mu,\nu=0, \ldots, D-1$ are spacetime indices and we use a mostly-minus metric. The $D$-dimensional gamma matrices obey the Clifford algebra relation
\begin{equation}
\{ \Gamma^\mu , \Gamma^\nu \} = 2 \eta^{\mu \nu} \ , 
\end{equation}
where all gamma matrices except $\Gamma^0$ are taken to be antihermitian (our conventions are summarized in Appendix \ref{appConv}).  $\hat a, \hat b, \hat c$ are adjoint indices of the gauge group.\footnote{For notational clarity, we will hat all gauge-group indices throughout the paper.} The fermions' indices are not displayed explicitly. Four-dimensional fermions carry three different indices: spacetime spinor indices, gauge-group representation indices and global/flavor indices. Fermions obey reality conditions of the form
\begin{equation}
\bar \lambda = \lambda^t C V \ , \label{RPRcondition}
\end{equation}
where $C$ is a matrix acting on spinor and flavor indices and $V$ is a matrix acting on the gauge-group representation indices which is taken to be unitary and antisymmetric. The representation matrices for the fermions obey
\begin{equation}
V T^{\hat a}_R V^\dagger = - (T^{\hat a}_R)^* \ ,
\end{equation} 
i.e. the representation $R$ is pseudo-real.
The spinor $\lambda$ 
includes $P$ flavors of irreducible $SO(D-1,1)$ spinors, which obey reality (R) or pseudo-reality (PR) conditions, according to the choice of $C$ in equation (\ref{RPRcondition}). Specifically, we have 
\begin{equation} 
{\rm R:} \ \ C =   {\cal C}_{D}   \ , 
\qquad \quad {\rm PR:} \ \ C =   {\cal C}_{D}  \Omega \ , \label{RPRcond}
\end{equation}
where ${\cal C}_{D}$  is the $SO(D-1,1)$ charge-conjugation matrix and
$\Omega$ is an antisymmetric real matrix acting on the flavor indices. 
The dimensions in which R and PR conditions are appropriate are listed in Table \ref{tab1} together with the corresponding flavor symmetry.
For $D=6,10$ (mod $8$), Weyl conditions are compatible with R and PR conditions. Since representations with opposite chiralities are inequivalent, we need to introduce 
two distinct integers $P$ and $\dot P$ giving the number of irreducible $D$-dimensional spinors for each chirality. Results of this paper are more naturally expressed in terms of the number of four-dimensional fermions $n_F$, which is also listed in Table \ref{tab1}.

\begin{table}[t]
	\centering
	\begin{tabular}{c@{\hskip 1cm}c@{\hskip 1cm}c@{\hskip 1cm}c}
		$D$ &    $n_F(D,P,\dot{P})$ & conditions & flavor group \\  
		\hline
		$5$& $P$ &  R &   $SO(P)$   \\ 
		$6$ & $P \! + \! \dot P$ & RW   & $SO(P)\! \! \times \! SO(\dot{P})$  \\
		$7$ & $2P$ & R & $SO(P)$  \\
		$8$ & $4P$ & R/W  & $U(P)$ \\
		$9$ & $8P$ & PR  & $USp(2P)$  \\
		$10$  & $8P \! + \! 8\dot P$ & PRW   & $USp(2P)\! \! \times \! USp(2\dot{P})$  \\
		$11$  & $16P$ & PR  & $USp(2P)$  \\
		$12$  & $16P$ & R/W &   $U(P)$  \\
		$k \!+\! 8$   &  $16 \, n_F(k,P,\dot{P}) $ & as for $k$ & as for $k$     \\
	\end{tabular}
	\caption{\small Parameters in the construction of $\cN=2$ homogeneous Maxwell-Einstein supergravities as double copies. 
		The second column gives the number $n_F$ of  irreducible spinors once the non-supersymmetric gauge-theory factor is reduced to $4D$.
		Reality (R), pseudo-reality (PR) or Weyl (W) conditions are used to obtain irreducible spinors in $D$ dimensions. \label{tab1}}
\end{table}

\bigskip
The left gauge theory entering the double-copy construction is  
a $\cN=2$ sYM theory with a single half-hypermultiplet 
transforming in the same pseudo-real representation $R$ which appears in the other gauge theory.
This choice is motivated by the fact 
that a pseudo-real half-hypermultiplet is CPT self-conjugate and hence does not need to be augmented to a full hypermultiplet to have a sensible theory. It is convenient to consider a sYM theory in $D=6$ spacetime dimensions with Lagrangian
\begin{equation}
{\cal L}_L = -{1 \over 4} F^{\hat a}_{\mu \nu} F^{\hat a \mu \nu} + {i } \bar \psi \Gamma^\mu D_\mu  \psi  + D_\mu  \bar \varphi D^{\mu} \varphi  + {i \over 2 } \bar \chi \Gamma^\mu  D_\mu  \chi + \sqrt{2} \bar \psi^{\hat a} \bar \varphi T_R^{\hat a} \chi + \sqrt{2}  \bar \chi  T_R^{\hat a} \varphi \psi^{\hat a} \ ,  
\end{equation}
where  the covariant derivatives are 
\begin{eqnarray}
(D_\mu \psi)^{\hat a} &=& \partial_\mu \psi^{\hat a}  + g f^{\hat a \hat b \hat c} A_\mu^{\hat b} \psi^{\hat c} \ , \\
D_\mu \chi &=& \partial_\mu \chi  - i g T_R^{\hat a} A_\mu^{\hat a} \chi \ , \\
D_\mu \varphi &=& \partial_\mu \varphi  - i g T_R^{\hat a} A_\mu^{\hat a} \varphi \ .
\end{eqnarray}
The spin-$1/2$ field $\psi^{\hat a}$ transforms in the adjoint representation and is the supersymmetric partner of the gluon. It obeys a chirality condition of the form
\begin{equation}
\Gamma_7  \psi = \psi \ , \qquad \Gamma_7 = \Gamma^0 \cdots \Gamma^5 \ .
\end{equation} 
The complex scalar $\varphi$ and the spinor $\chi$ are the components of the half-hypermultiplet. They obey the conditions,
\begin{equation}
\bar \chi  = \chi^t C V \ , \qquad \ \Gamma_7  \chi = \chi   \ . 
\end{equation} 
Amplitudes in this theory can  be conveniently organized into superamplitudes 
with manifest ${\cal N}=2$ supersymmetry. While the Lagrangian in this section can be used for Feynman-rule computations, the quickest way to obtain amplitudes for the left theory is to start from $\cN=4$ amplitudes and use an orbifold procedure, as explained in the next section. 

Finally, a generic homogeneous supergravity in four-dimensions has a U-duality algebra which can be decomposed into grade $0,1,2$ generators, with the grade-zero part
\begin{equation}
 {\cal G}_0 = so(1,1) \oplus so( D-4 ,2) \oplus {\cal S}(P, \dot P) \ , 
\end{equation}
where ${\cal S}(P, \dot P)$ is the flavor algebra listed in Table \ref{tab1}. A $so(D-4) \oplus {\cal S}(P, \dot P)$ subalgebra is already manifest at the level of the gauge theories entering the construction. \\

\section{One-loop gauge amplitudes with hypermultiplets \label{sec3}}
\renewcommand{\theequation}{3.\arabic{equation}}
\setcounter{equation}{0}

Field-theory orbifolds \cite{Bershadsky:1998cb} are constructed from a parent theory, which in this case is $\cN=4$ sYM, by projecting out states that are not invariant under the transformation
\begin{equation}
\Phi \rightarrow R_i g_i \Phi g_i^\dagger \ , \qquad (R_i,g_i) \in \Gamma \ ,
\end{equation}
where $\Phi$ is a generic field of the theory, $g_i$ is an element of the gauge group $G$ and $R_i$ is a matrix acting on the R-symmetry indices of the theory. $\Gamma$ is taken to be a discrete abelian subgroup of $G \times SU(4)$. Orbifold amplitudes can be conveniently obtained by inserting the projectors 
\begin{equation}
 {\cal P}_\Gamma \Phi = {1 \over |\Gamma|} \sum_{(R_i,g_i) \in \Gamma} R_i g_i \Phi g_i^\dagger \ ,
\end{equation} 
in the propagators of the parent theory, where $|\Gamma|$ is the order of the orbifold group. As a consequence of the invariance of the Feynman-rule vertices under gauge and R-symmetry, projectors can be moved around the diagram past individual vertices. For tree-level amplitudes, this observation implies that orbifold amplitudes are identical to the ones of the parent theory, provided that the external states are chosen to be invariant under the orbifold group.  
One-loop amplitudes that preserve color/kinematics duality for general abelian orbifolds of $\cN=4$ sYM theory were obtained in ref. \cite{Chiodaroli:2013upa}. They are generally organized in a presentation based on cubic graphs in which the internal line labeled by the loop momentum is dressed with an extra phase factor, which is in turn expressed in terms of the entries of a diagonal matrix encoding the action of the R-symmetry part of the orbifold generators on fundamental $SU(4)$ indices.    

Without repeating the derivation in ref. \cite{Chiodaroli:2013upa} and referring to Appendix  \ref{Apporbifold} for details, we focus  on amplitudes with four external hypermultiplets. 
Hypermultiplet asymptotic states can be conveniently organized in an on-shell superfield ${\cal Q}$,
\begin{equation}
{\cal Q} = \chi_{+} + \eta^r \varphi_{r} + \eta^1 \eta^2 \tilde \chi_{-} \ , \quad r=1,2 \ ,
\end{equation}
together with the CPT-conjugate superfield $\overline{\cal Q}$. When the representation is pseudo-real, we can identify ${\cal Q}=\overline{\cal Q}$ and 
$\chi_\pm = \tilde \chi_\pm$.
The superamplitude between four hypermultiplets is then written as
\begin{equation}
{\cal F}_4^{1-\text{loop}} \big( 1 {\cal Q} , 2 {\cal Q}, 3 \overline{ \cal Q}, 4 \overline{\cal Q } \big)
= - g^{4}\sum_{{\cal G}_3} {1 \over S_i} {n_i C_i \over D_i} \label{ampscal}  \ .
\end{equation}
Color and numerator factors are listed in Table \ref{complexrep}. While we have obtained these amplitudes with a particular choice of gauge group and representation matrices, amplitudes with any other choice of gauge group and complex representation have the same formal expression. In later sections, we will simply use the numerators without any 
explicit reference to the orbifold procedure we have employed in their derivation.\\

It is interesting to look at the color and numerator relations obeyed by the above amplitude. 
Some color relations stem from the representation-matrices commutation relations $[\tilde T^{\hat a}, \tilde T^{\hat b}]= \tilde f^{\hat a \hat b \hat c } \tilde T^{\hat c}$:\footnote{Throughout the paper we have $\tilde f^{\hat a \hat b \hat c} = i \sqrt{2} f^{\hat a \hat b \hat c}$ and $\tilde T^{\hat a} =  \sqrt{2} T^{\hat a}$. }
\begin{eqnarray}
C_1 - C_{2b} &=& C_{5b} = C_{7b} \ , \no \\
C_3 - C_{2a} &=&  C_{8b} = - C_{9b}  \ , \no  \end{eqnarray}
\begin{eqnarray}	
C_{11a} &=& 2 C_{8b} = - 2 C_{9b} \ , \no  \\
C_{12a} &=& 2 C_{5b} = 2 C_{7b} \ . \label{color-rel-basic}
\end{eqnarray}
The reader can verify that these relations are obeyed by the corresponding numerator factors.\\

\def\widthdiag{0.2\textwidth}
\def\widthone{0.05\textwidth}
\def\widththree{0.65\textwidth}

\begin{longtable}{c|c|c} \small
			$i$ & ${\cal G}$ & $C_i, n_i$ \\
			\hline
			\parbox{\widthone}{$1$} &
			\parbox[c]{\widthdiag}{\ \\ \includegraphics[width=\widthdiag]{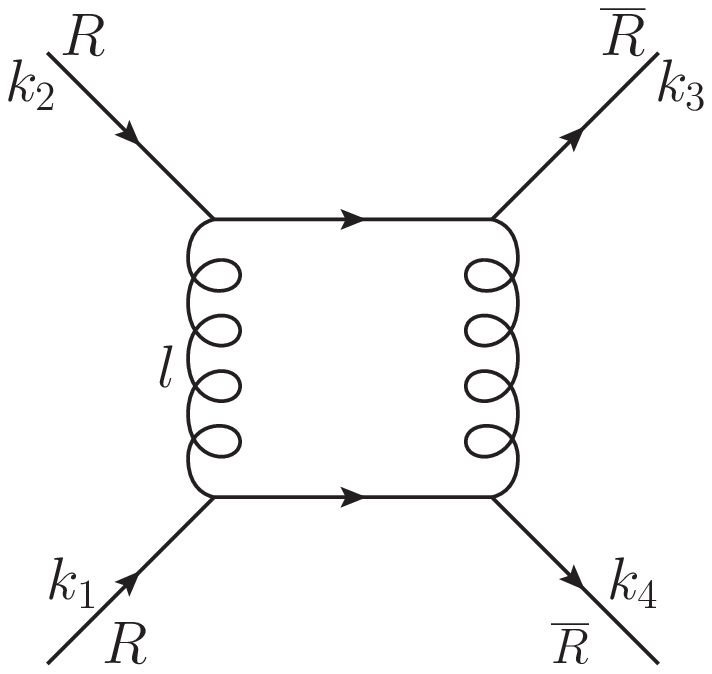} \\  } & 
			\parbox{\widththree}{
				\begin{eqnarray}
				C_1 & = & (\tilde T^{\hat a}\tilde T^{\hat b})_{\hat \alpha}^{\ {\hat \delta}} (\tilde T^{\hat a}\tilde T^{\hat b})_{\hat \beta}^{\ {\hat \gamma}} \no \\
				n_1 & = & {s^2 \over \langle 12 \rangle \langle 34 \rangle} \delta^4 \big( Q \big) 
				\end{eqnarray} } \\[5pt]
			\hline
			\parbox{\widthone}{$2a$} &
			\parbox{\widthdiag}{\ \\ \includegraphics[width=\widthdiag]{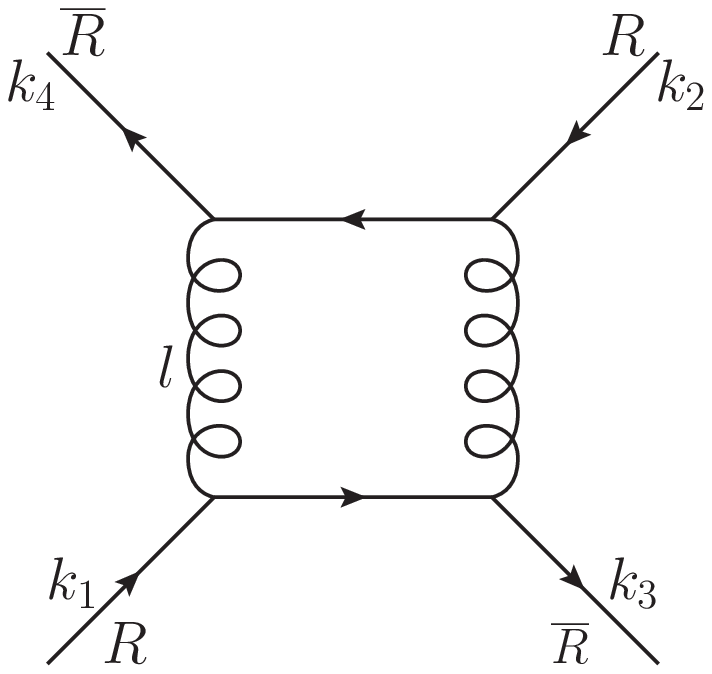} \\ } & 
			\parbox{\widththree}{
				\begin{eqnarray}
				C_{2a} & = & (\tilde T^{\hat a}\tilde T^{\hat b})_{\hat \alpha}^{\ {\hat \gamma}} (\tilde T^{\hat b}\tilde T^{\hat a})_{\hat \beta}^{\ {\hat \delta}} \no \\
				n_{2a} & = & - {st \over \langle 12 \rangle \langle 34 \rangle} \delta^4 \big( Q \big)
				\end{eqnarray} } \\[5pt]
			\hline
			\parbox{\widthone}{$2b$} &
			\parbox{\widthdiag}{\ \\ \includegraphics[width=\widthdiag]{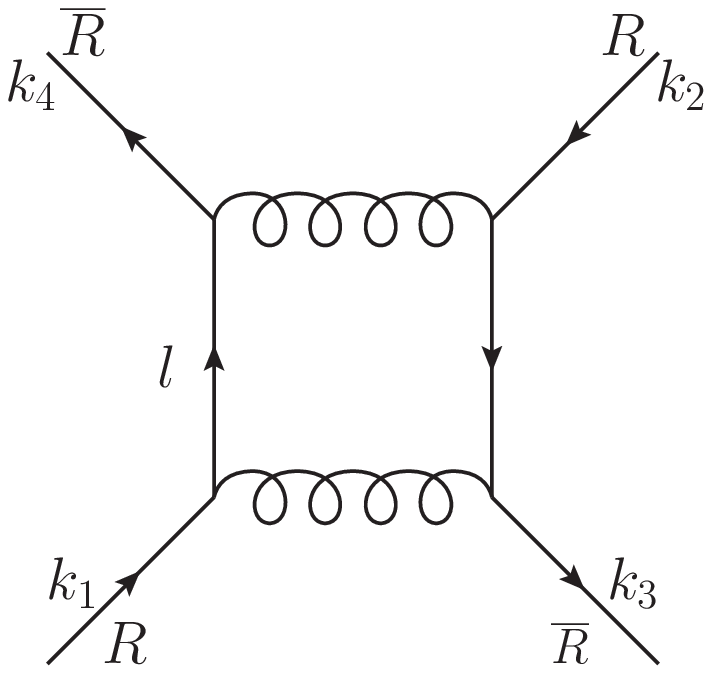} \\ } & 
			\parbox{\widththree}{
				\begin{eqnarray}
				C_{2b} & = & (\tilde T^{\hat a}\tilde T^{\hat b})_{\hat \alpha}^{\ {\hat \delta}} (\tilde T^{\hat b}\tilde T^{\hat a})_{\hat \beta}^{\ {\hat \gamma}} \no \\
				n_{2b} & = & - {s u \over \langle 12 \rangle \langle 34 \rangle} \delta^4 \big( Q \big)
				\end{eqnarray} } \\[5pt]
			\hline
			\parbox{\widthone}{$3$} &
			\parbox{\widthdiag}{\ \\ \includegraphics[width=\widthdiag]{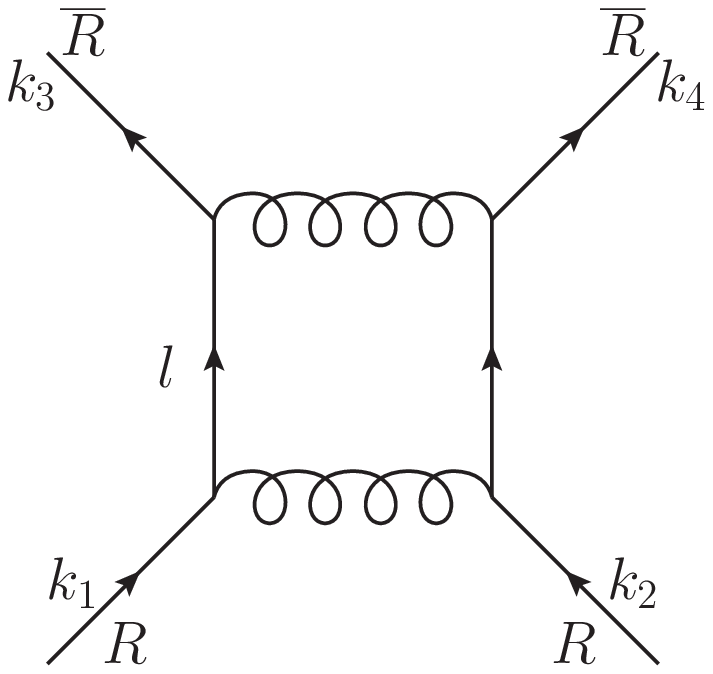} \\ } & 
			\parbox{\widththree}{
				\begin{eqnarray}
				C_{3} & = & (\tilde T^{\hat a}\tilde T^{\hat b})_{\hat \alpha}^{\ {\hat \gamma}} (\tilde T^{\hat a}\tilde T^{\hat b})_{\hat \beta}^{\ {\hat \delta}} \no \\
				n_{3} & = &  {s^2 \over \langle 12 \rangle \langle 34 \rangle} \delta^4 \big( Q \big)
				\end{eqnarray} } \\[5pt]
			\hline
		\parbox{\widthone}{$5a$} &
		\parbox{\widthdiag}{\includegraphics[width=\widthdiag]{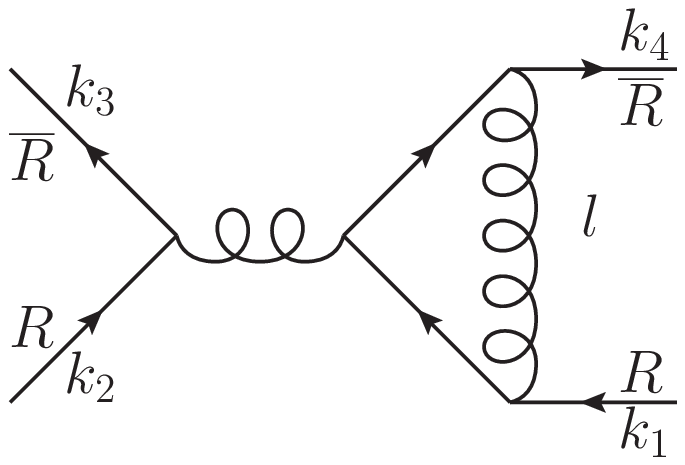}} & 
		\parbox{\widththree}{
			\begin{eqnarray}
			C_{5a} & = & - \tilde T^{\hat a \ \hat \gamma }_{\ \hat  \beta} (\tilde T^{\hat b}\tilde T^{\hat a}\tilde T^{\hat b})_{\hat \alpha}^{\ {\hat \delta}} = \Big( T(G)
			- 2C(R)\Big)  \tilde T^{\hat a \ \hat \delta }_{\ \hat  \alpha} \tilde T^{\hat a \ \hat \gamma }_{\ \hat  \beta} \no \\
			n_{5a} & = &  {st \over \langle 12 \rangle \langle 34 \rangle} \delta^4 \big(Q \big)
			\end{eqnarray} } \\[5pt]
		\hline
		\parbox{\widthone}{$5b$} &
		\parbox{\widthdiag}{\includegraphics[width=\widthdiag]{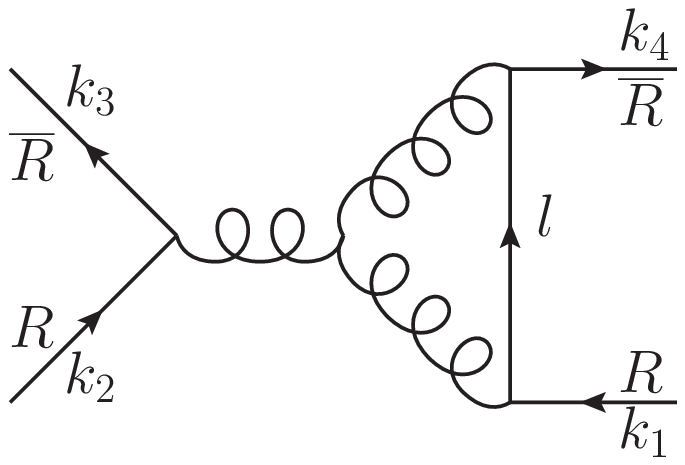}} & 
		\parbox{\widththree}{
			\begin{eqnarray}
			C_{5b} & = &  \tilde T^{\hat a \ \hat \gamma }_{\ \hat \beta}  \tilde f^{\hat a \hat b \hat c} (\tilde T^{\hat b}\tilde T^{\hat c})_{\hat \alpha}^{\ {\hat \delta}} \no =  -  T(G) \tilde T^{\hat a \ \hat \delta }_{\ \hat  \alpha} \tilde T^{\hat a \ \hat \gamma }_{\ \hat \beta} \no \\
			n_{5b} & = & -  {st \over \langle 12 \rangle \langle 34 \rangle} \delta^4 \big( Q \big)
			\end{eqnarray} } \\[5pt]
		\hline
		\parbox{\widthone}{$7a$} &
		\parbox{\widthdiag}{\includegraphics[width=\widthdiag]{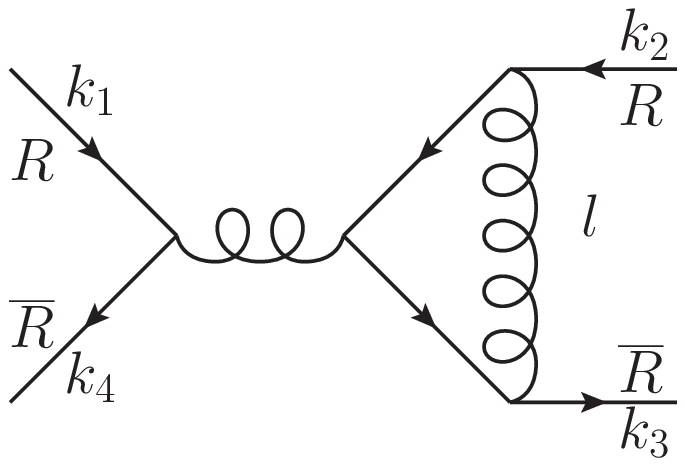}} & 
		\parbox{\widththree}{
			\begin{eqnarray}
			C_{7a} & = & - \tilde T^{\hat a \ \hat \delta }_{\ \hat  \alpha} (\tilde T^{\hat b}\tilde T^{\hat a}\tilde T^{\hat b})_{\hat \beta}^{\ {\hat \gamma}} = \Big( T(G)
			- 2 C(R)\Big) \tilde T^{\hat a \ \hat \delta }_{\ \hat  \alpha} \tilde T^{\hat a \ \hat \gamma }_{\ \hat  \beta}  \no \\
			n_{7a} & = &   {st \over \langle 12 \rangle \langle 34 \rangle} \delta^4 \big( Q \big)
			\end{eqnarray} } \\[5pt]
		\hline
		\parbox{\widthone}{$7b$} &
		\parbox{\widthdiag}{\includegraphics[width=\widthdiag]{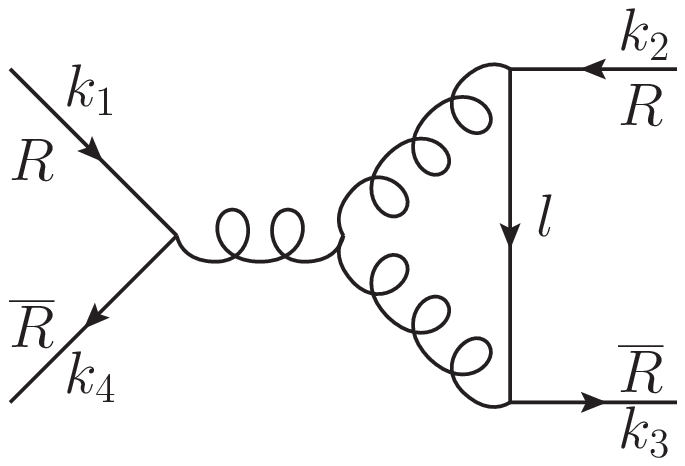}} & 
		\parbox{\widththree}{
			\begin{eqnarray}
			C_{7b} & = &  \tilde T^{\hat a \ \hat \delta }_{\ \hat  \alpha}  \tilde f^{\hat a \hat b \hat c} (\tilde T^{\hat b}\tilde T^{\hat c})_{\hat \beta}^{\ {\hat \gamma}} = - T(G)  \tilde T^{\hat a \ \hat \delta }_{\ \hat  \alpha} \tilde T^{\hat a \ \hat \gamma }_{\ \hat  \beta} \no \\
			n_{7b} & = & - {s t \over \langle 12 \rangle \langle 34 \rangle} \delta^4 \big( Q \big)
			\end{eqnarray} } \\[5pt]
		\hline
		\parbox{\widthone}{$8a$} &
		\parbox{\widthdiag}{\includegraphics[width=\widthdiag]{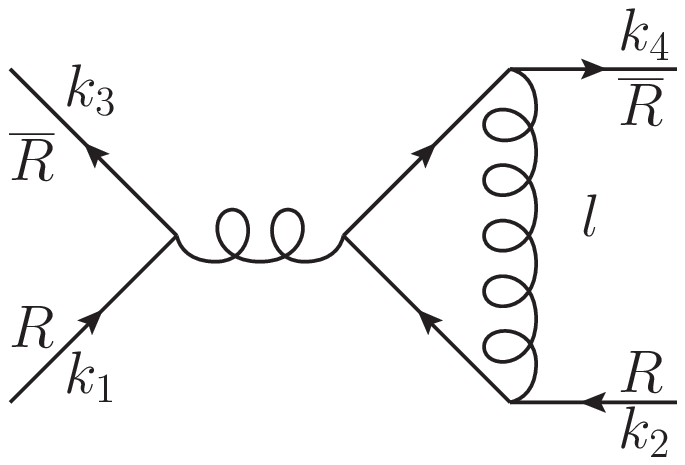}} & 
		\parbox{\widththree}{
			\begin{eqnarray}
			C_{8a} & = & - \tilde T^{\hat a \ \hat \gamma }_{\ \hat  \alpha} (\tilde T^{\hat b}\tilde T^{\hat a}\tilde T^{\hat b})_{\hat \beta}^{\ {\hat \delta}} = \Big( T(G)
			- 2 C(R)\Big) \tilde T^{\hat a \ \hat \gamma }_{\ \hat  \alpha}  \tilde T^{\hat a \ \hat \delta }_{\ \hat  \beta} \no \\
			n_{8a} & = &  {s u \over \langle 12 \rangle \langle 34 \rangle} \delta^4 \big( Q \big)
			\end{eqnarray} } \\[5pt]
		\hline
		\parbox{\widthone}{$8b$} &
		\parbox{\widthdiag}{\includegraphics[width=\widthdiag]{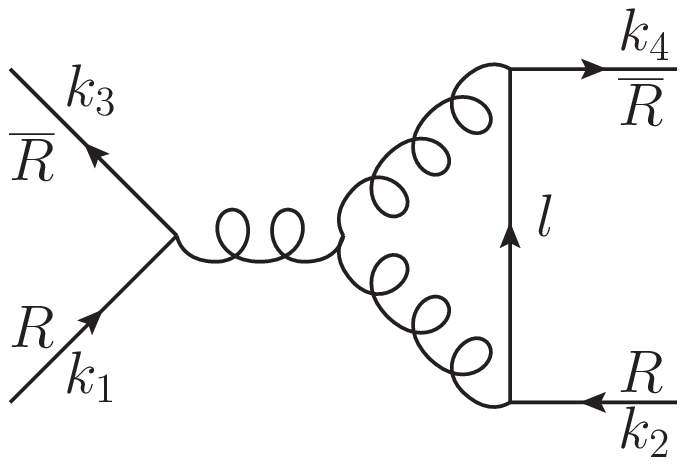}} & 
		\parbox{\widththree}{
			\begin{eqnarray}
			C_{8b} & = &  \tilde T^{\hat a \ \hat \gamma }_{\ \hat  \alpha} \tilde f^{\hat a \hat b \hat c} (\tilde T^{\hat b}\tilde T^{\hat c})_{\hat \beta}^{\ {\hat \delta}}= - T(G) \tilde T^{\hat a \ \hat \gamma }_{\ \hat  \alpha} \tilde T^{\hat a \ \hat \delta }_{\ \hat  \beta} \no \\
			n_{8b} & = & - {s u \over \langle 12 \rangle \langle 34 \rangle} \delta^4 \big( Q \big)
			\end{eqnarray} } \\[5pt]
		\hline
		\parbox{\widthone}{$9a$} &
		\parbox{\widthdiag}{\includegraphics[width=\widthdiag]{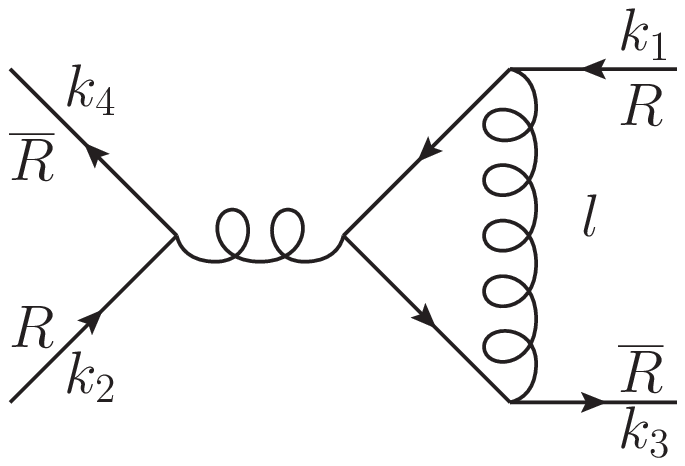}} & 
		\parbox{\widththree}{
			\begin{eqnarray}
			C_{9a} & = &  \tilde T^{\hat a \ \hat \delta }_{\ \hat  \beta} (\tilde T^{\hat b}\tilde T^{\hat a}\tilde T^{\hat b})_{\hat \alpha}^{\ {\hat \gamma}} = - \Big(  T(G)
			- 2 C(R)\Big) \tilde T^{\hat a \ \hat \gamma }_{\ \hat  \alpha} \tilde T^{\hat a \ \hat \delta }_{\ \hat  \beta} \no \\
			n_{9a} & = & - {s u \over \langle 12 \rangle \langle 34 \rangle} \delta^4 \big( Q \big)
			\end{eqnarray} } \\[5pt]
		\hline
		\parbox{\widthone}{$9b$} &
		\parbox{\widthdiag}{\includegraphics[width=\widthdiag]{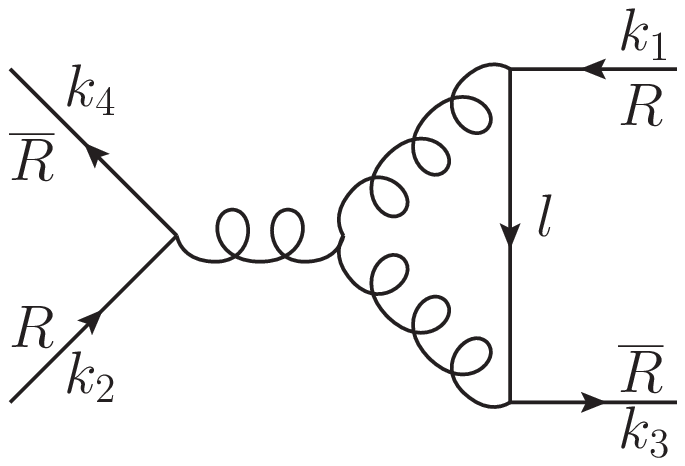}} & 
		\parbox{\widththree}{
			\begin{eqnarray}
			C_{9b} & = & - \tilde T^{\hat a \ \hat \delta }_{\ \hat  \beta}  \tilde f^{\hat a \hat b \hat c} (\tilde T^{\hat b}\tilde T^{\hat c})_{\hat \alpha}^{\ {\hat \gamma}} =    T(G) \tilde T^{\hat a \ \hat \gamma }_{\ \hat  \alpha} \tilde T^{\hat a \ \hat \delta }_{\ \hat  \beta} \no \\
			n_{9b} & = &   {s u \over \langle 12 \rangle \langle 34 \rangle} \delta^4 \big( Q \big)
			\end{eqnarray} } \\[5pt]
		\hline
			\parbox{\widthone}{$11a$} &
		\parbox{\widthdiag}{\includegraphics[width=\widthdiag]{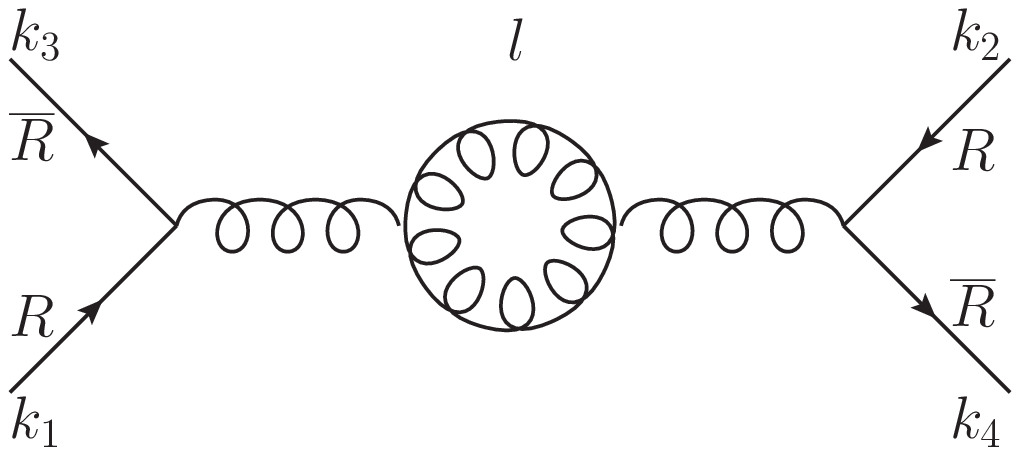}} & 
		\parbox{\widththree}{
			\begin{eqnarray}
			C_{11a} & = &  \tilde T^{\hat a \ \hat \gamma }_{\ \hat  \alpha} \tilde f^{\hat a \hat b \hat c} \tilde f^{\hat b \hat c \hat d} \tilde T^{\hat a \ \hat \delta }_{\ \hat  \beta} =  - 2 T(G) \tilde T^{\hat a \ \hat \gamma }_{\ \hat  \alpha} \tilde T^{\hat a \ \hat \delta }_{\ \hat  \beta} \no \\
			n_{11a} & = & - 2 {s u \over \langle 12 \rangle \langle 34 \rangle} \delta^4 \big( Q \big)
			\end{eqnarray} } \\[5pt]
		\hline
		\parbox{\widthone}{$11b$} &
		\parbox{\widthdiag}{\includegraphics[width=\widthdiag]{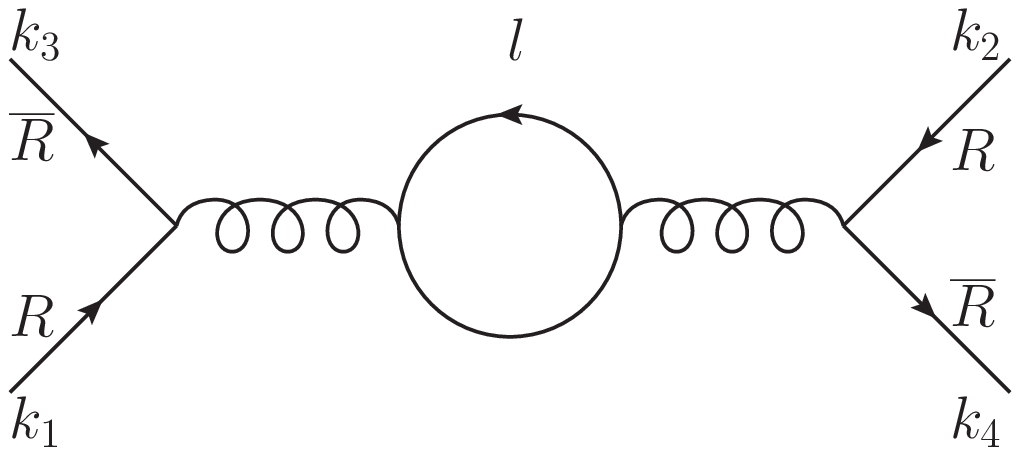}} & 
		\parbox{\widththree}{
			\begin{eqnarray}
			C_{11b} & = & -\tilde T^{\hat a \ \hat \gamma }_{\ \hat  \alpha} {\rm Tr }(\tilde T^{\hat a} \tilde T^{\hat b}) \tilde T^{\hat a \ \hat \delta }_{\ \hat  \beta} = - 2 T(R) \tilde T^{\hat a \ \hat \gamma }_{\ \hat  \alpha} \tilde T^{\hat a \ \hat \delta }_{\ \hat  \beta} \no \\
			n_{11b} & = &   {s u \over \langle 12 \rangle \langle 34 \rangle} \delta^4 \big( Q \big)
			\end{eqnarray} } \\[5pt]
		\hline
		\parbox{\widthone}{$12a$} &
		\parbox{\widthdiag}{\includegraphics[width=\widthdiag]{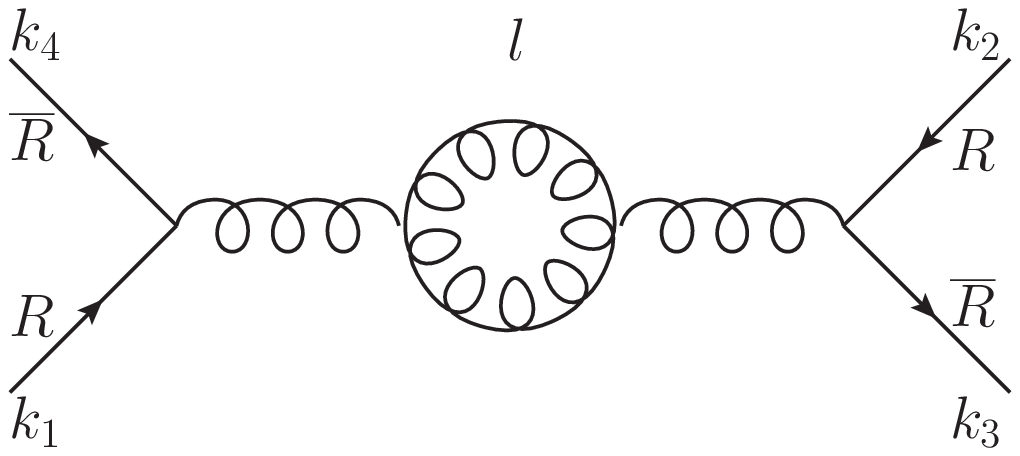}} & 
		\parbox{\widththree}{
			\begin{eqnarray}
			C_{12a} & = &  \tilde T^{\hat a \ \hat \delta }_{\ \hat  \alpha} \tilde f^{\hat a \hat b \hat c} \tilde f^{\hat b \hat c \hat d} \tilde T^{\hat a \ \hat \gamma }_{\ \hat  \beta}  = - 2 T(G) \tilde T^{\hat a \ \hat \delta }_{\ \hat  \alpha} \tilde T^{\hat a \ \hat \gamma }_{\ \hat  \beta} \no \\
			n_{12a} & = & - 2  {s t \over \langle 12 \rangle \langle 34 \rangle} \delta^4 \big( Q \big)
			\end{eqnarray} } \\[5pt]
		\hline
		\parbox{\widthone}{$12b$} &
		\parbox{\widthdiag}{\includegraphics[width=\widthdiag]{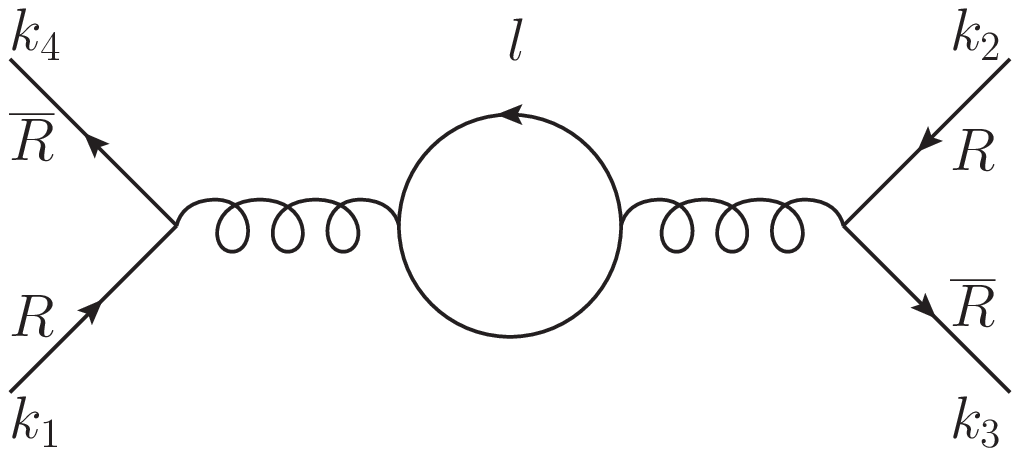}} & 
		\parbox{\widththree}{
			\begin{eqnarray}
			C_{12b} & = & - \tilde T^{\hat a \ \hat \delta }_{\ \hat  \alpha} {\rm Tr }(\tilde T^{\hat a} \tilde T^{\hat b}) \tilde T^{\hat a \ \hat \gamma }_{\ \hat  \beta}  =  - 2 T(R) \tilde T^{\hat a \ \hat \delta }_{\ \hat  \alpha} \tilde T^{\hat a \ \hat \gamma }_{\ \hat  \beta} \no \\
			n_{12b} & = &  {s t \over \langle 12 \rangle \langle 34 \rangle} \delta^4 \big( Q \big)
			\end{eqnarray} } \\[5pt]
\hline \multicolumn{3}{c}{} \\
	\caption{Color and numerator factors for the 1-loop superamplitude with four external hypermultiplets in case of a complex representation\label{complexrep}. Curly lines denote adjoint fields, while solid lines denote hypermultiplet matter fields.  Arrows are assigned according to the representation and we use the short-hand notation $\delta^4 \big( Q \big) = \delta^4 \big( \sum_i \eta_i^r |i\rangle \big)$. The numbering of the graphs follows the one in ref. \cite{Chiodaroli:2013upa}. Gauge-group representation indices $\hat \alpha, \hat \beta$ are written explicitly. }
\end{longtable}

Other relations stem from the two-term identity
$T^{\hat a \ \hat \gamma}_{\ \hat \alpha} T^{\hat a \ \hat \delta}_{\ \hat \beta} = T^{\hat a \ \hat \delta}_{\ \hat \alpha} T^{\hat a \ \hat \gamma}_{\ \hat \beta}$:
\begin{eqnarray}
C_1 &=& C_3 \ , \no \\
C_{2b} &=& - C_{8a} =  C_{9a} \ ,\no \\
C_{2a} &=&  - C_{5a} = - C_{7a}  \ , \no \\
C_{11b} &=&  C_{8a} = - C_{9a} \ , \no \\
C_{12b} &=&  C_{7a} =  C_{5a} \ . \label{additional}
\end{eqnarray}
These relations are also obeyed by the corresponding numerator factors. If we chose a different complex representation with respect to the one from orbifolding, the color relations would no longer hold, but numerator relations would be unaltered. 
For later convenience, color factors in Table \ref{complexrep} are rewritten in terms of the representation's index $T(R)$ and quadratic Casimir $C(R)$,
\begin{equation}
\Tr \big( \tilde T_R^{\hat a} \tilde T_R^{\hat b}   \big) = 2 T(R) \delta^{\hat a \hat b} \ ,\qquad  \tilde T^{\hat a}_R \tilde T^{\hat a}_R = 2 C(R) \ {\mathbf{1} }_R \ .
\end{equation}

While the numerators obtained so far are appropriate for complex representations, the construction outlined in the previous section notably uses pseudo-real representations as a key ingredient. It is quite instructive to see how the construction is modified in this case. On general grounds, using a reality condition allows us to relate fermions with their Dirac conjugates. For four-fermion amplitudes, this implies that a (fermionic) permutation symmetry needs to be introduced on all external legs. This symmetry extends to superamplitudes with four-external half-hypermultiplets, taking into account that hypermultiplet on-shell superfields anticommute with each other (since their lowest component is Grassmann-odd). In short, duality-satisfying numerators can be written down for the pseudo-real case as the unique set of numerators with the following two properties:
\begin{enumerate}
	\item Numerators are invariant under the permutation of all external legs up to a sign which is assigned according to the signature of the permutation.
	\item Numerators corresponding to color factors which are non-zero in the case of a complex representation reproduce the ones listed in the Table \ref{complexrep}.  
\end{enumerate} 
The complete list of numerators is given in Table \ref{prealrep}.\\

\begin{longtable}[c]{c|c|c} \small
	$i$ & ${\cal G}$ & $C_i, n_i$ \\
	\hline
	\parbox{\widthone}{$1a$} &
	\parbox[c]{\widthdiag}{\ \\ \includegraphics[width=\widthdiag]{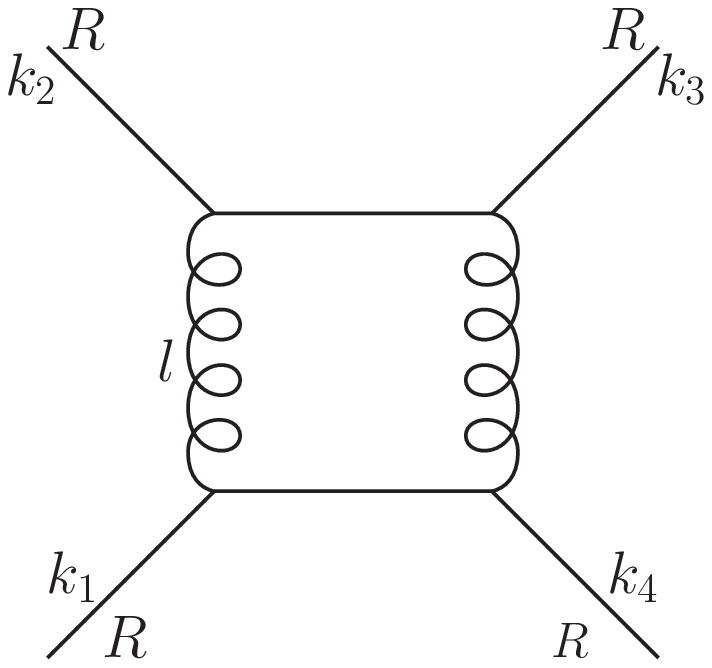} \\  } & 
	\parbox{\widththree}{
		\begin{eqnarray}
		C_{1a} & = & (\tilde T^{\hat a}\tilde T^{\hat b})_{\hat \alpha {\hat \delta}} (\tilde T^{\hat a}\tilde T^{\hat b})_{\hat \beta {\hat \gamma}} \no \\
		n_{1a} & = & {s^2 \over \langle 12 \rangle \langle 34 \rangle} \delta^4 \big( Q \big) 
		\end{eqnarray} } \\[5pt]
	\hline
		\parbox{\widthone}{$1b$} &
		\parbox[c]{\widthdiag}{\ \\ \includegraphics[width=\widthdiag]{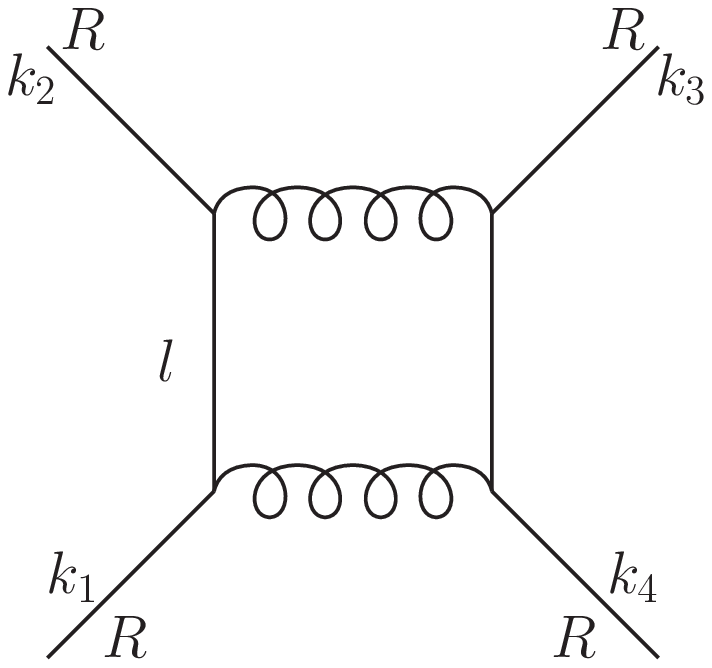} \\  } & 
		\parbox{\widththree}{
			\begin{eqnarray}
			C_{1b} & = & (\tilde T^{\hat a}\tilde T^{\hat b})_{\hat \alpha {\hat \beta}} (\tilde T^{\hat b}\tilde T^{\hat a})_{\hat \gamma {\hat \delta}} \no \\
			n_{1b} & = & -{st \over \langle 12 \rangle \langle 34 \rangle} \delta^4 \big( Q \big) 
			\end{eqnarray} } \\[5pt]
		\hline
	\parbox{\widthone}{$2a$} &
	\parbox{\widthdiag}{\ \\ \includegraphics[width=\widthdiag]{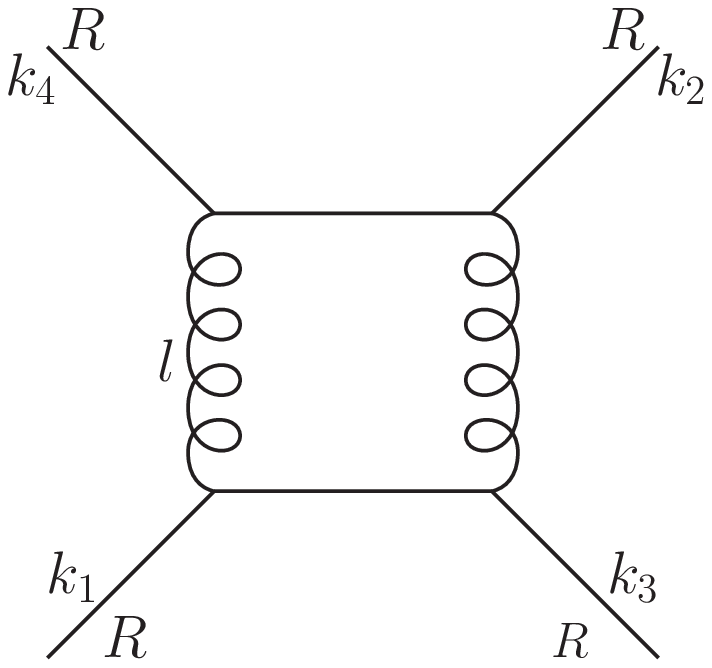} \\ } & 
	\parbox{\widththree}{
		\begin{eqnarray}
		C_{2a} & = & (\tilde T^{\hat a}\tilde T^{\hat b})_{\hat \alpha {\hat \gamma}} (\tilde T^{\hat b}\tilde T^{\hat a})_{\hat \beta {\hat \delta}} \no \\
		n_{2a} & = & - {st \over \langle 12 \rangle \langle 34 \rangle} \delta^4 \big( Q \big)
		\end{eqnarray} } \\[5pt]
	\hline
	\parbox{\widthone}{$2b$} &
	\parbox{\widthdiag}{\  \includegraphics[width=\widthdiag]{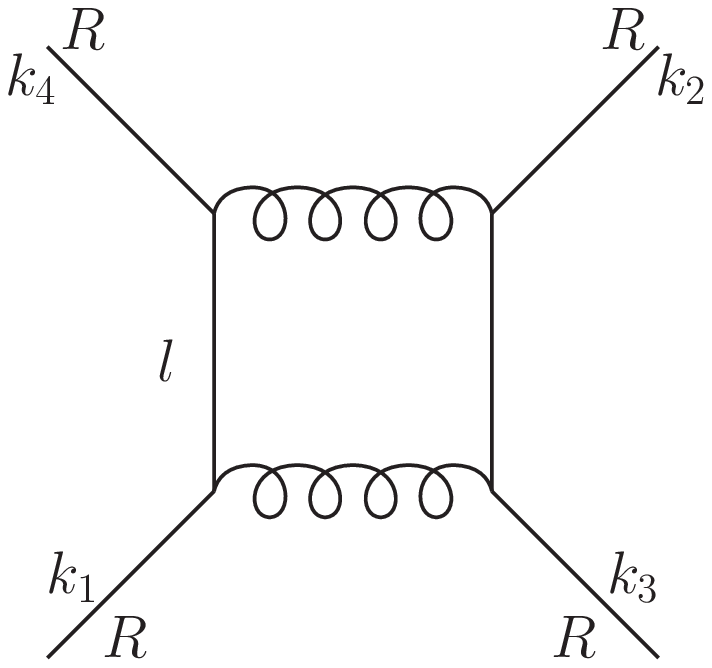} \\ } & 
	\parbox{\widththree}{
		\begin{eqnarray}
		C_{2b} & = & (\tilde T^{\hat a}\tilde T^{\hat b})_{\hat \alpha {\hat \delta}} (\tilde T^{\hat b}\tilde T^{\hat a})_{\hat \beta {\hat \gamma}} \no \\
		n_{2b} & = & - {s u \over \langle 12 \rangle \langle 34 \rangle} \delta^4 \big( Q \big)
		\end{eqnarray} } \\[5pt]
	\hline
	\parbox{\widthone}{$3a$} &
	\parbox{\widthdiag}{\  \includegraphics[width=0.2\textwidth]{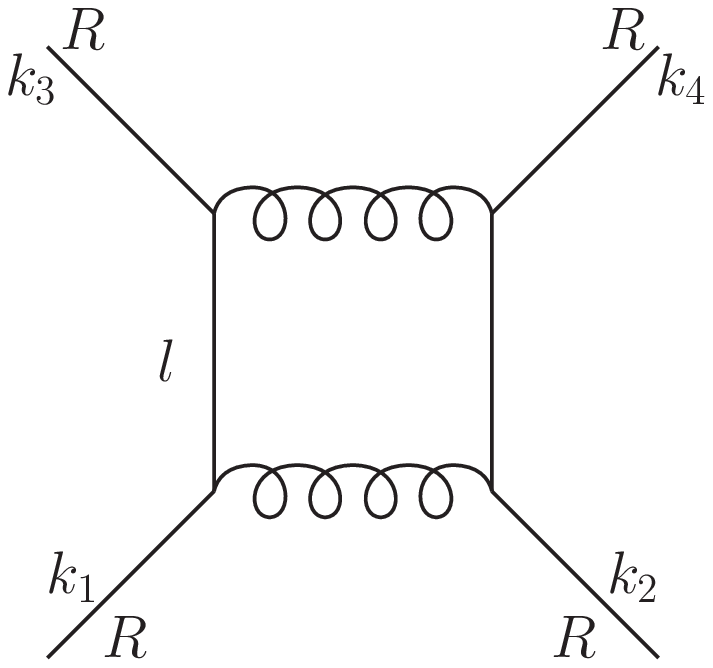} \\ } & 
	\parbox{\widththree}{
		\begin{eqnarray}
		C_{3a} & = & (\tilde T^{\hat a}\tilde T^{\hat b})_{\hat \alpha {\hat \gamma}} (\tilde T^{\hat a}\tilde T^{\hat b})_{\hat \beta {\hat \delta}} \no \\
		n_{3a} & = &  {s^2 \over \langle 12 \rangle \langle 34 \rangle} \delta^4 \big( Q \big)
		\end{eqnarray} } \\[5pt]
	\hline
	\parbox{\widthone}{$3b$} &
	\parbox{\widthdiag}{\  \includegraphics[width=\widthdiag]{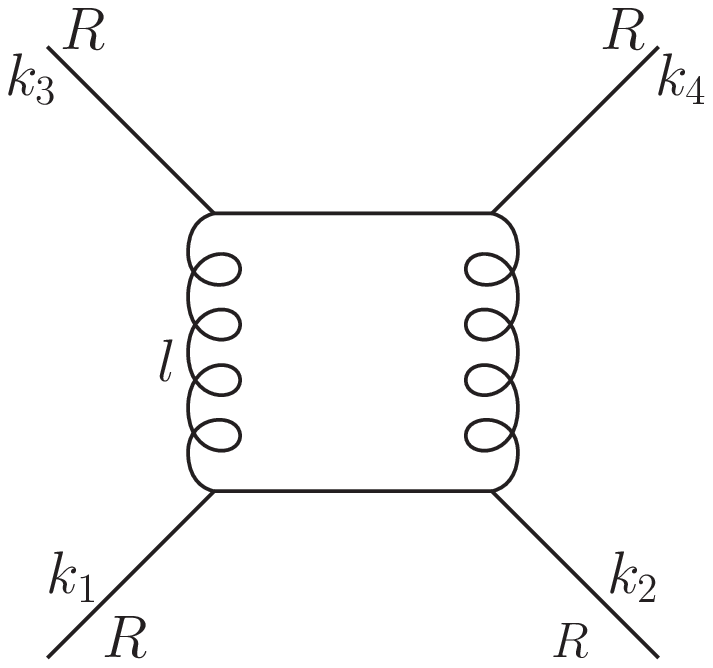} \\ } & 
	\parbox{\widththree}{
		\begin{eqnarray}
		C_{3b} & = & (\tilde T^{\hat a}\tilde T^{\hat b})_{\hat \alpha  {\hat \beta}} (\tilde T^{\hat a}\tilde T^{\hat b})_{\hat \gamma {\hat \delta}} \no \\
		n_{3b} & = &  {s u \over \langle 12 \rangle \langle 34 \rangle} \delta^4 \big( Q \big)
		\end{eqnarray} } \\[5pt]
	\hline
	\parbox{\widthone}{$4a$} &
	\parbox{\widthdiag}{\includegraphics[width=\widthdiag]{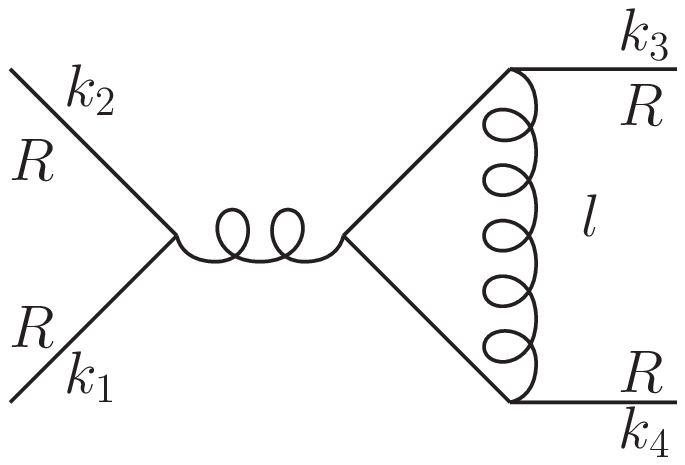}} & 
	\parbox{\widththree}{
		\begin{eqnarray}
		C_{4a} & = & - \tilde T^{\hat a}_{\hat \alpha {\hat \beta}} (\tilde T^{\hat b}\tilde T^{\hat a}\tilde T^{\hat b})_{\hat \gamma {\hat \delta}} = \Big( T(G)
		- 2C(R)\Big) \tilde T^{\hat a}_{ \hat \alpha \hat \beta} \tilde T^{\hat a}_{ \hat \gamma \hat \delta} \no \\
		n_{4a} & = &  {s^2 \over \langle 12 \rangle \langle 34 \rangle} \delta^4 \big(Q \big)
		\end{eqnarray} } \\[5pt]
	\hline
	\parbox{\widthone}{$4b$} &
	\parbox{\widthdiag}{\includegraphics[width=\widthdiag]{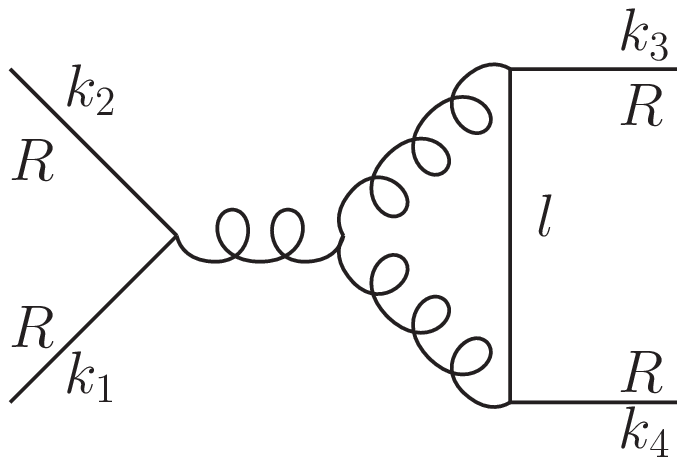}} & 
	\parbox{\widththree}{
		\begin{eqnarray}
		C_{4b} & = &  \tilde T^{\hat a}_{\hat \alpha {\hat \beta}}  \tilde f^{\hat a \hat b \hat c} (\tilde T^{\hat b}\tilde T^{\hat c})_{ \hat \gamma {\hat \delta}} \no =  -  T(G) \tilde T^{\hat a  }_{ \hat \alpha  \hat \beta} \tilde T^{\hat a}_{\hat \gamma  \hat \delta  } \no \\
		n_{4b} & = & - {s^2 \over \langle 12 \rangle \langle 34 \rangle} \delta^4 \big( Q \big)
		\end{eqnarray} } \\[5pt]
	\hline
	\parbox{\widthone}{$5a$} &
	\parbox{\widthdiag}{\includegraphics[width=\widthdiag]{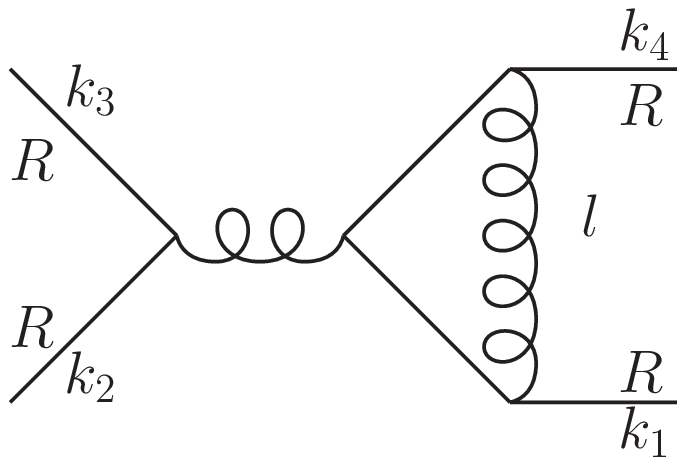}} & 
	\parbox{\widththree}{
		\begin{eqnarray}
		C_{5a} & = & - \tilde T^{\hat a}_{\hat \beta {\hat \gamma}} (\tilde T^{\hat b}\tilde T^{\hat a}\tilde T^{\hat b})_{\hat \alpha {\hat \delta}} = \Big( T(G)
		- 2C(R)\Big) \tilde T^{\hat a}_{\hat \beta  \hat \gamma} \tilde T^{\hat a  }_{\hat \alpha \delta} \no \\
		n_{5a} & = &  {st \over \langle 12 \rangle \langle 34 \rangle} \delta^4 \big(Q \big)
		\end{eqnarray} } \\[5pt]
	\hline
	\parbox{\widthone}{$5b$} &
	\parbox{\widthdiag}{\includegraphics[width=\widthdiag]{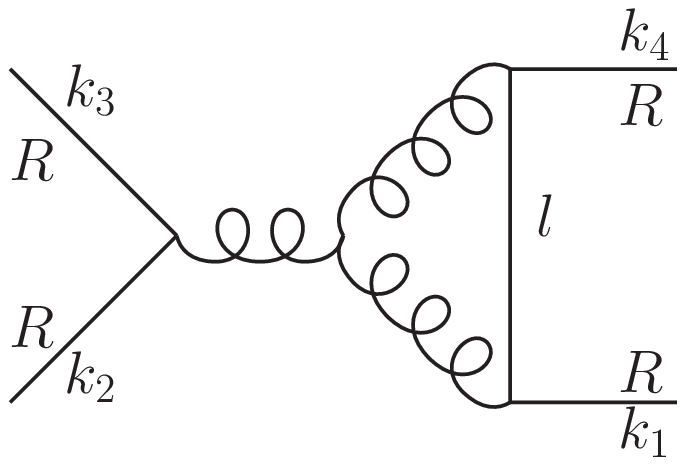}} & 
	\parbox{\widththree}{
		\begin{eqnarray}
		C_{5b} & = &  \tilde T^{\hat a}_{\hat \beta {\hat \gamma}}  \tilde f^{\hat a \hat b \hat c} (\tilde T^{\hat b}\tilde T^{\hat c})_{\hat \alpha {\hat \delta}} \no =  -  T(G) \tilde T^{\hat a}_{\hat \beta \hat \gamma } \tilde T^{\hat a }_{\hat \alpha \hat \delta} \no \\
		n_{5b} & = & -  {st \over \langle 12 \rangle \langle 34 \rangle} \delta^4 \big( Q \big)
		\end{eqnarray} } \\[5pt]
	\hline
		\parbox{\widthone}{$6a$} &
		\parbox{\widthdiag}{\includegraphics[width=\widthdiag]{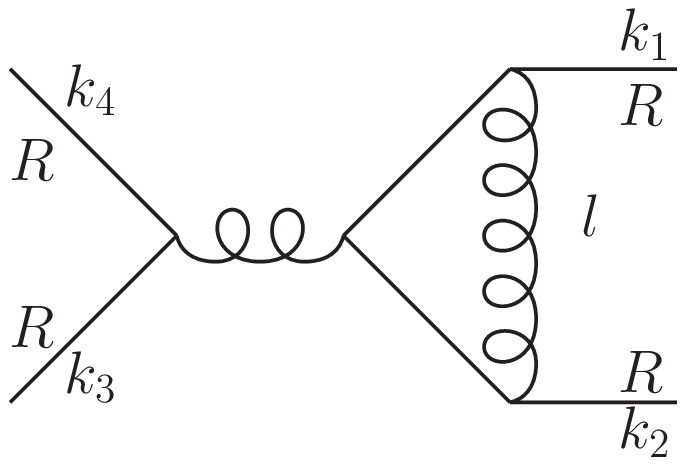}} & 
		\parbox{\widththree}{
			\begin{eqnarray}
			C_{6a} & = & - \tilde T^{\hat a}_{\hat \gamma {\hat \delta}} (\tilde T^{\hat b}\tilde T^{\hat a}\tilde T^{\hat b})_{\hat \alpha {\hat \beta}} = \Big( T(G)
			- 2 C(R)\Big) \tilde T^{\hat a}_{\hat \alpha \hat \beta} \tilde T^{\hat a}_{\hat \gamma  \hat \delta } \no \\
			n_{6a} & = &   {s^2 \over \langle 12 \rangle \langle 34 \rangle} \delta^4 \big( Q \big)
			\end{eqnarray} } \\[5pt]
		\hline
		\parbox{\widthone}{$6b$} &
		\parbox{\widthdiag}{\includegraphics[width=\widthdiag]{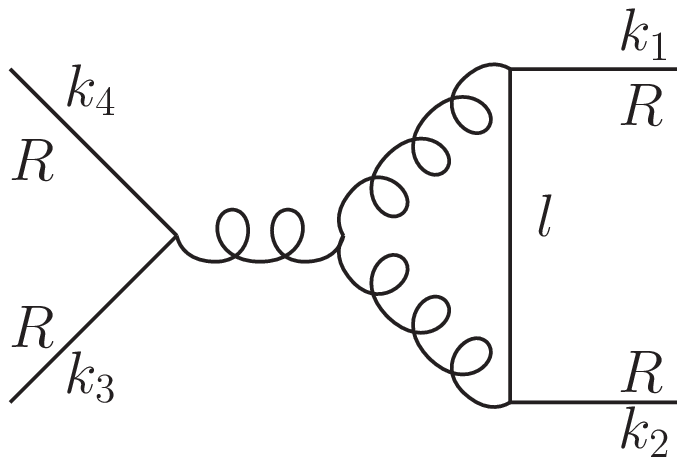}} & 
		\parbox{\widththree}{
			\begin{eqnarray}
			C_{6b} & = &  \tilde T^{\hat a}_{\hat \gamma {\hat \delta}}  \tilde f^{\hat a \hat b \hat c} (\tilde T^{\hat b}\tilde T^{\hat c})_{\hat \alpha {\hat \beta}} = - T(G) \tilde T^{\hat a}_{ \hat \alpha \hat \beta} \tilde T^{\hat a }_{\hat \gamma  \hat \delta } \no \\
			n_{6b} & = &  -{s^2  \over \langle 12 \rangle \langle 34 \rangle} \delta^4 \big( Q \big)
			\end{eqnarray} } \\[5pt]
		\hline

	\parbox{\widthone}{$7a$} &
	\parbox{\widthdiag}{\includegraphics[width=\widthdiag]{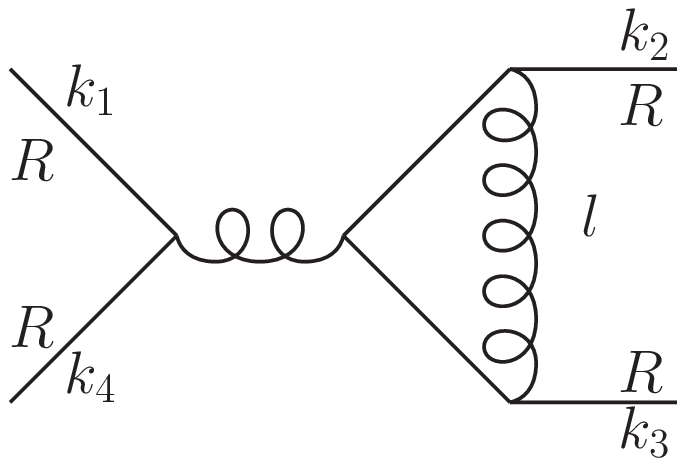}} & 
	\parbox{\widththree}{
		\begin{eqnarray}
		C_{7a} & = & - \tilde T^{\hat a}_{\hat \alpha {\hat \delta}} (\tilde T^{\hat b}\tilde T^{\hat a}\tilde T^{\hat b})_{\hat \beta {\hat \gamma}} = \Big( T(G)
		- 2 C(R)\Big) \tilde T^{\hat a}_{\hat \beta \hat \gamma} \tilde T^{\hat a}_{\hat \alpha  \hat \delta } \no \\
		n_{7a} & = &   {st \over \langle 12 \rangle \langle 34 \rangle} \delta^4 \big( Q \big)
		\end{eqnarray} } \\[5pt]
	\hline
	\parbox{\widthone}{$7b$} &
	\parbox{\widthdiag}{\includegraphics[width=\widthdiag]{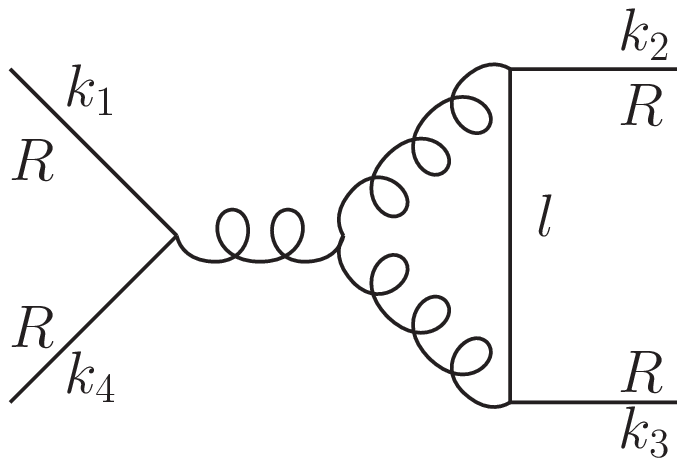}} & 
	\parbox{\widththree}{
		\begin{eqnarray}
		C_{7b} & = &  \tilde T^{\hat a}_{\hat \alpha {\hat \delta}}  \tilde f^{\hat a \hat b \hat c} (\tilde T^{\hat b}\tilde T^{\hat c})_{\hat \beta {\hat \gamma}} = - T(G) \tilde T^{\hat a}_{ \hat \beta \hat \gamma} \tilde T^{\hat a }_{\hat \alpha  \hat \delta } \no \\
		n_{7b} & = & - {s t \over \langle 12 \rangle \langle 34 \rangle} \delta^4 \big( Q \big)
		\end{eqnarray} } \\[5pt]
	\hline
	\parbox{\widthone}{$8a$} &
	\parbox{\widthdiag}{\includegraphics[width=\widthdiag]{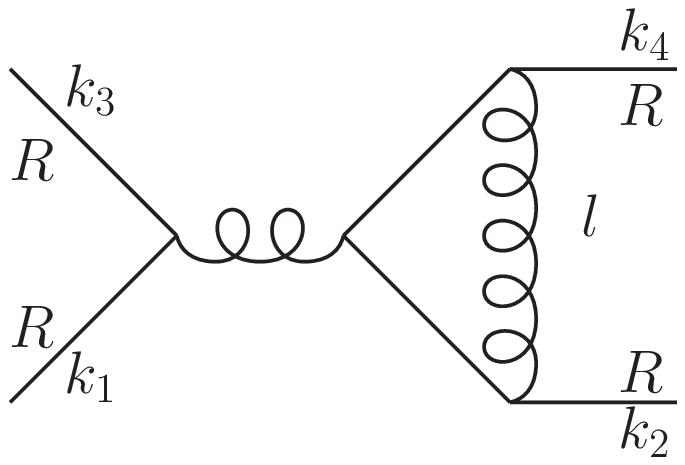}} & 
	\parbox{\widththree}{
		\begin{eqnarray}
		C_{8a} & = & -\tilde T^{\hat a}_{\hat \alpha {\hat \gamma}} (\tilde T^{\hat b}\tilde T^{\hat a}\tilde T^{\hat b})_{\hat \beta  {\hat \delta}} = \Big( T(G)
		- 2 C(R)\Big) \tilde T^{\hat a }_{\hat \alpha \hat \gamma}  \tilde T^{\hat a}_{ \hat \beta \hat \delta} \no \\
		n_{8a} & = &  {s u \over \langle 12 \rangle \langle 34 \rangle} \delta^4 \big( Q \big)
		\end{eqnarray} } \\[5pt]
	\hline
	\parbox{\widthone}{$8b$} &
	\parbox{\widthdiag}{\includegraphics[width=\widthdiag]{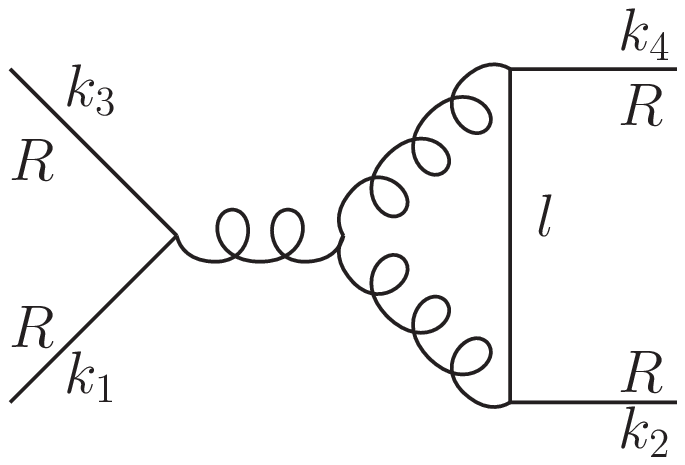}} & 
	\parbox{\widththree}{
		\begin{eqnarray}
		C_{8b} & = &  \tilde T^{\hat a}_{\hat \alpha {\hat \gamma}}  \tilde f^{\hat a \hat b \hat c} (\tilde T^{\hat b}\tilde T^{\hat c})_{\hat \beta {\hat \delta}}= - T(G) \tilde T^{\hat a}_{\hat \alpha \hat \gamma} \tilde T^{\hat a}_{\hat \beta  \hat \delta } \no \\
		n_{8b} & = & - {s u \over \langle 12 \rangle \langle 34 \rangle} \delta^4 \big( Q \big)
		\end{eqnarray} } \\[5pt]
	\hline
	\parbox{\widthone}{$9a$} &
	\parbox{\widthdiag}{\includegraphics[width=\widthdiag]{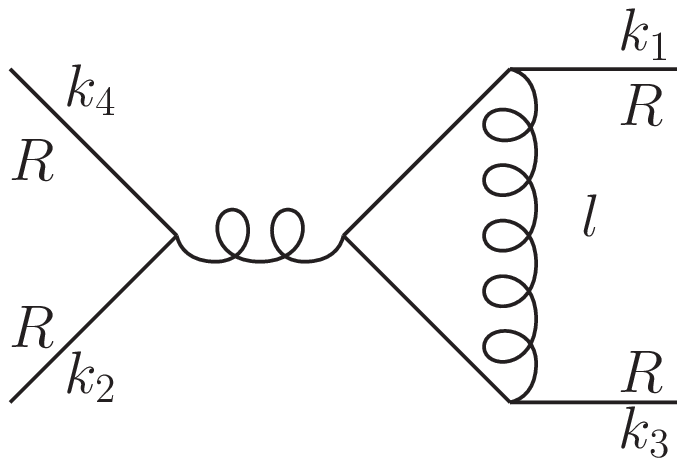}} & 
	\parbox{\widththree}{
		\begin{eqnarray}
		C_{9a} & = &  \tilde T^{\hat a}_{\hat \beta {\hat \delta}} (\tilde T^{\hat b}\tilde T^{\hat a}\tilde T^{\hat b})_{\hat \alpha {\hat \gamma}} = - \Big(  T(G)
		- 2 C(R)\Big) \tilde T^{\hat a}_{\hat \alpha \hat \gamma } \tilde T^{\hat a}_{\hat \beta  \hat \delta} \no \\
		n_{9a} & = & - {s u \over \langle 12 \rangle \langle 34 \rangle} \delta^4 \big( Q \big)
		\end{eqnarray} } \\[5pt]
	\hline
	\parbox{\widthone}{$9b$} &
	\parbox{\widthdiag}{\includegraphics[width=\widthdiag]{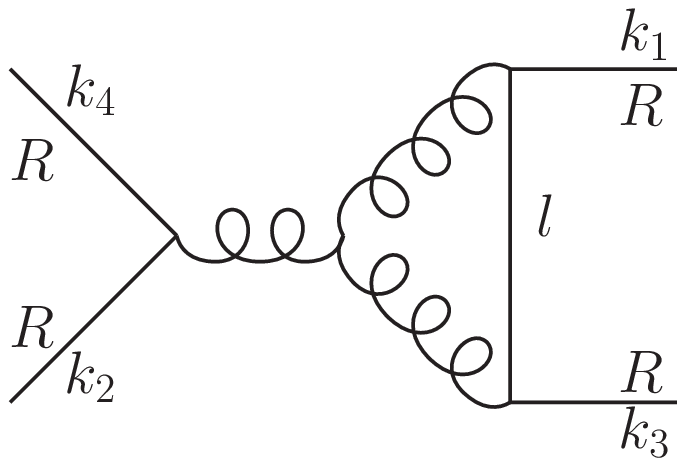}} & 
	\parbox{\widththree}{
		\begin{eqnarray}
		C_{9b} & = & -\tilde T^{\hat a}_{\hat \beta {\hat \delta}}  \tilde f^{\hat a \hat b \hat c} (\tilde T^{\hat b}\tilde T^{\hat c})_{\hat \alpha {\hat \gamma}} =    T(G) \tilde T^{\hat a}_{\hat \alpha \hat \gamma} \tilde T^{\hat a}_{\hat \beta \hat \delta} \no \\
		n_{9b} & = &   {s u \over \langle 12 \rangle \langle 34 \rangle} \delta^4 \big( Q \big)
		\end{eqnarray} } \\[5pt]
		\hline
		\parbox{\widthone}{$10a$} &
		\parbox{\widthdiag}{\includegraphics[width=\widthdiag]{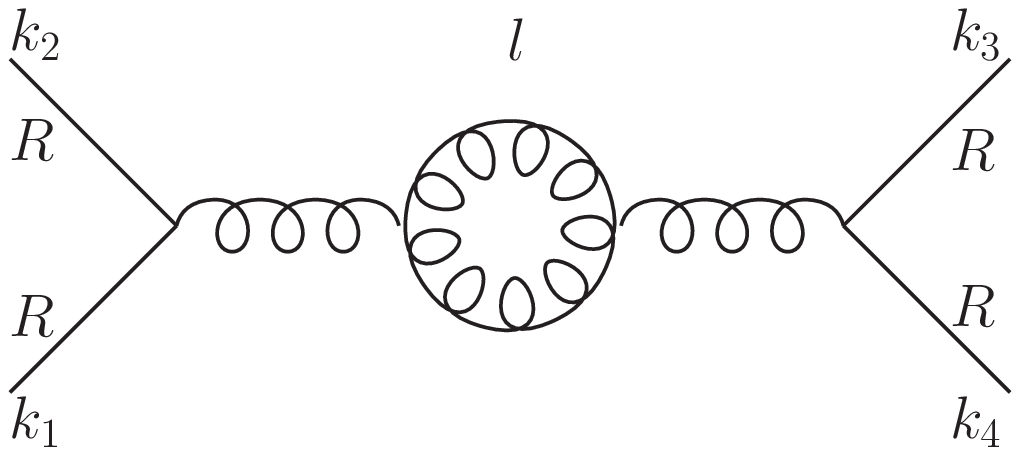}} & 
		\parbox{\widththree}{
			\begin{eqnarray}
			C_{10a} & = &  \tilde T^{\hat a}_{\hat \alpha {\hat \beta}} \tilde f^{\hat a \hat b \hat c} \tilde f^{\hat b \hat c \hat d} \tilde T^d_{\hat \gamma {\hat \delta}} =  - 2  T(G) \tilde T^{\hat a }_{\hat \alpha \hat \beta} \tilde T^{\hat a }_{\hat \gamma  \hat \delta } \no \\
			n_{10a} & = &  - 2 {s^2  \over \langle 12 \rangle \langle 34 \rangle} \delta^4 \big( Q \big)
			\end{eqnarray} } \\[5pt]
		\hline
		\parbox{\widthone}{$10b$} &
		\parbox{\widthdiag}{\includegraphics[width=\widthdiag]{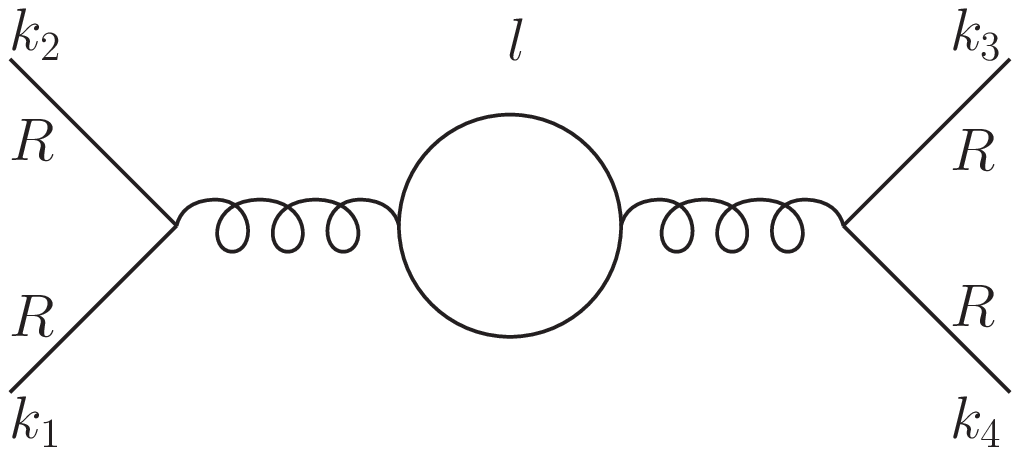}} & 
		\parbox{\widththree}{
			\begin{eqnarray}
			C_{10b} & = & -\tilde T^{\hat a}_{\hat \alpha {\hat \beta}} {\rm Tr }(\tilde T^{\hat a} \tilde T^{\hat b}) \tilde T^{\hat b}_{\hat \gamma {\hat \delta}} = - 2 T(R) \tilde T^{\hat a}_{\hat \alpha  \hat \beta } \tilde T^{\hat a  }_{\hat \gamma  \hat \delta} \no \\
			n_{10b} & = &   {s^2  \over \langle 12 \rangle \langle 34 \rangle} \delta^4 \big( Q \big)
			\end{eqnarray} } \\[5pt]	
	\hline
	\parbox{\widthone}{$11a$} &
	\parbox{\widthdiag}{\includegraphics[width=\widthdiag]{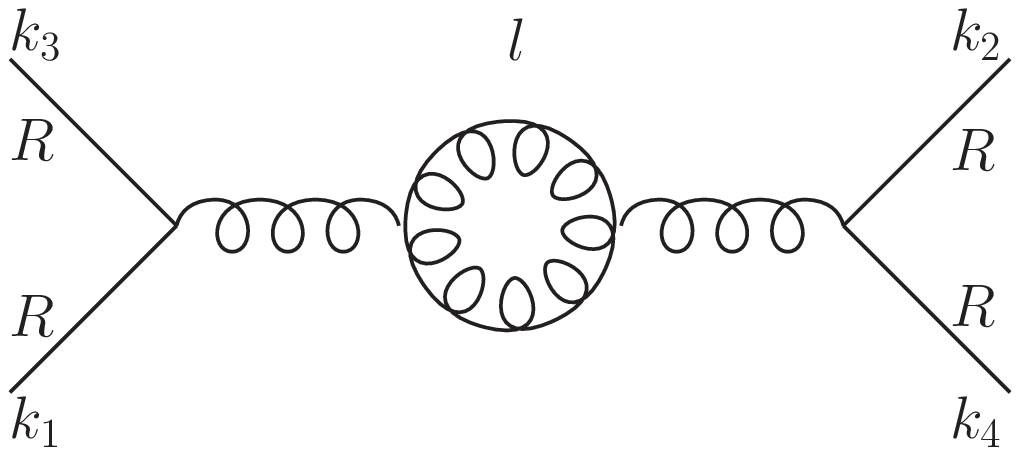}} & 
	\parbox{\widththree}{
		\begin{eqnarray}
		C_{11a} & = &  \tilde T^{\hat a}_{\hat \alpha {\hat \gamma}} \tilde f^{\hat a \hat b \hat c} \tilde f^{\hat b \hat c \hat d} \tilde T^d_{\hat \beta {\hat \delta}} =  -  2 T(G) \tilde T^{\hat a }_{\hat \alpha \hat \gamma} \tilde T^{\hat a }_{\hat \beta  \hat \delta } \no \\
		n_{11a} & = & - 2 {s u \over \langle 12 \rangle \langle 34 \rangle} \delta^4 \big( Q \big)
		\end{eqnarray} } \\[5pt]
	\hline
	\parbox{\widthone}{$11b$} &
	\parbox{\widthdiag}{\includegraphics[width=\widthdiag]{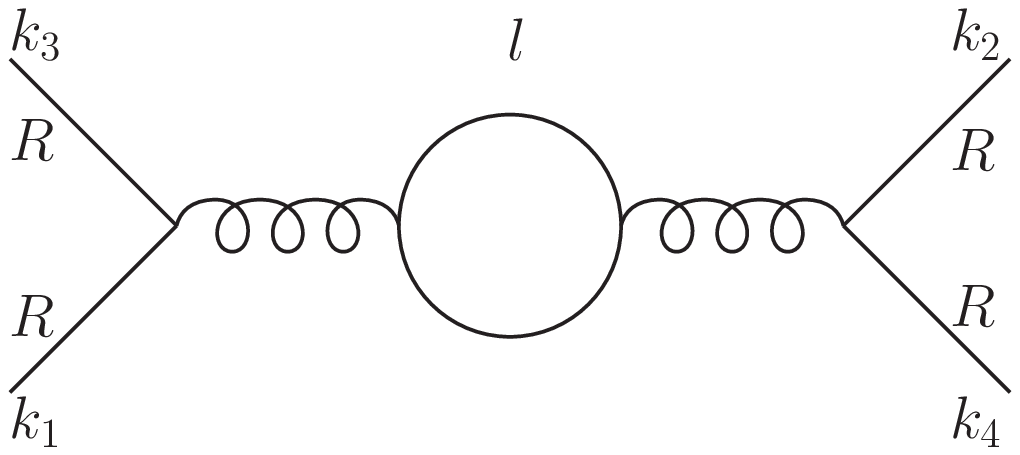}} & 
	\parbox{\widththree}{
		\begin{eqnarray}
		C_{11b} & = & -\tilde T^{\hat a}_{\hat \alpha {\hat \gamma}} {\rm Tr }(\tilde T^{\hat a} \tilde T^{\hat b}) \tilde T^{\hat b}_{\hat \beta {\hat \delta}} = - 2 T(R) \tilde T^{\hat a}_{\hat \alpha  \hat \gamma } \tilde T^{\hat a  }_{\hat \beta  \hat \delta} \no \\
		n_{11b} & = &   {s u \over \langle 12 \rangle \langle 34 \rangle} \delta^4 \big( Q \big)
		\end{eqnarray} } \\[5pt]
	\hline
	\parbox{\widthone}{$12a$} &
	\parbox{\widthdiag}{\includegraphics[width=\widthdiag]{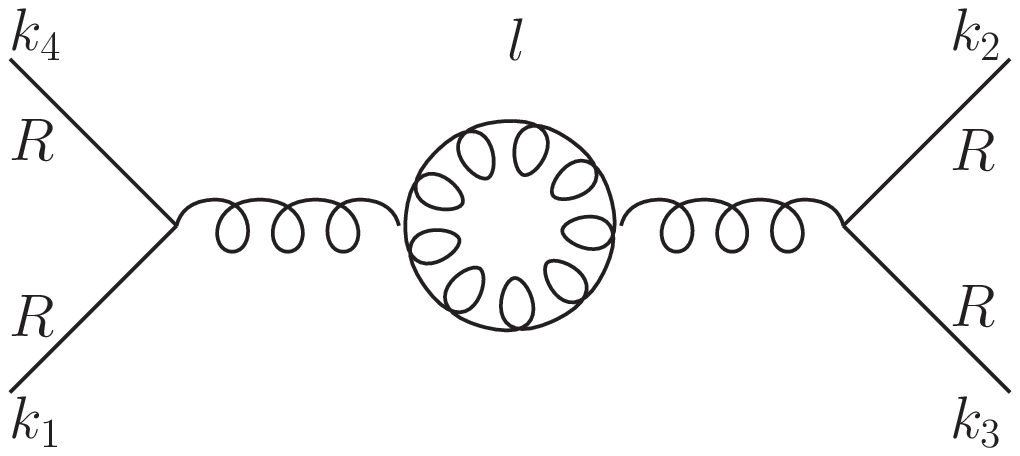}} & 
	\parbox{\widththree}{
		\begin{eqnarray}
		C_{12a} & = &  \tilde T^{\hat a}_{\hat \alpha {\hat \delta}} \tilde f^{\hat a \hat b \hat c} \tilde f^{\hat b \hat c \hat d} \tilde T^d_{\hat \beta {\hat \gamma}} =  - 2 T(G) \tilde T^{\hat a}_{\hat \alpha  \hat \delta } \tilde T^{\hat a }_{\hat \beta \hat \gamma} \no \\
		n_{12a} & = & - 2  {s t \over \langle 12 \rangle \langle 34 \rangle} \delta^4 \big( Q \big)
		\end{eqnarray} } \\[5pt]
	\hline
	\parbox{\widthone}{$12b$} &
	\parbox{\widthdiag}{\includegraphics[width=\widthdiag]{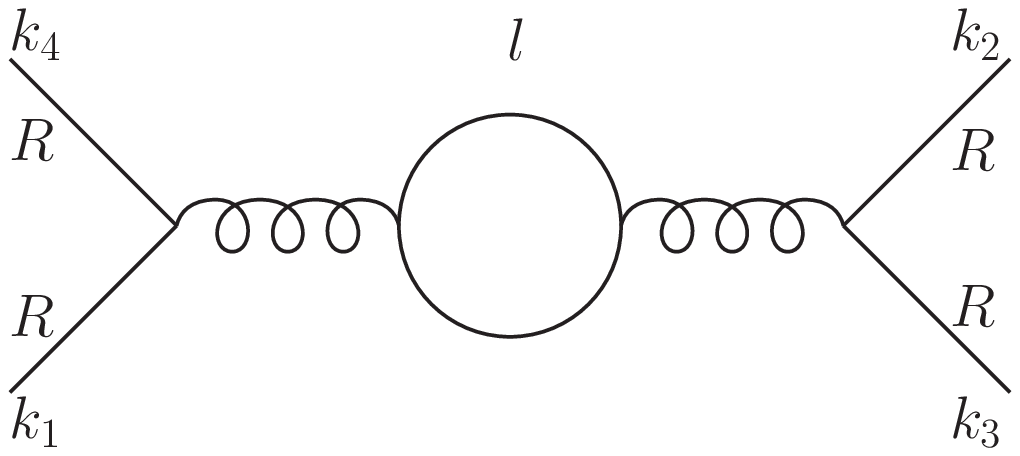}} & 
	\parbox{\widththree}{
		\begin{eqnarray}
		C_{12b} & = & - \tilde T^{\hat a}_{\hat \alpha {\hat \delta}} {\rm Tr }(\tilde T^{\hat a} \tilde T^{\hat b}) \tilde T^{\hat b}_{\hat \beta {\hat \gamma}} =  - 2 T(R) \tilde T^{\hat a }_{\hat \alpha  \hat \delta} \tilde T^{\hat a  }_{\hat \beta \hat \gamma} \no \\
		n_{12b} & = &  {s t \over \langle 12 \rangle \langle 34 \rangle} \delta^4 \big( Q \big)
		\end{eqnarray} } \\[5pt]
	\hline \multicolumn{3}{c}{} \\
	\caption{Color and numerator factors for the 1-loop superamplitude with four external half-hypermultiplets in case of a pseudo-real representation. Note that all bubbles have symmetry factor $S_i=2$. Representation matrices with two low indices are defined as $T^{\hat a}_{\hat \alpha \hat \beta} = (T^{\hat a} V^{-1})_{\hat \alpha \hat \beta}$, where $V$ is the antisymmetric matrix entering the pseudo-reality condition. \label{prealrep}}
\end{longtable}
Numerators with different assignments of external legs are related 
by permutation symmetry, as it can be verified using the identities
\begin{equation}
{s \over \langle 12\rangle \langle 34\rangle} = - {u \over \langle 13\rangle \langle 24\rangle} = {t \over \langle 14\rangle \langle 23\rangle} \ .
\end{equation}
The identity 
\begin{equation}
(\tilde T^{\hat a}  \tilde T^{\hat b})^t = - \tilde T^{\hat b}  \tilde T^{\hat a} \ ,
\end{equation}
which relies on the antisymmetry of $V$, is also useful.
An important difference between amplitudes in the complex and pseudo-real cases is that the symmetry factor for the bubble integrals with internal half-hypermultiplets needs to be changed to take into account that the corresponding graphs no longer carry arrows. 

The additional nonzero color and numerator factors obey identities of the form
\begin{eqnarray}
C_{1b} - C_{3b} &=&  - C_{6b} = - C_{4b} \no \\
C_{4b} + C_{6b} &=& C_{10a} 
\end{eqnarray}
which stem from the gauge-group generators commutation relations.
However, the extra identities in which gauge-group generators are contracted by an adjoint index require a more detailed discussion. 
In contrast to the complex case, color factors will no longer obey identities of the form $T^{\hat a \ \hat \gamma}_{\ \hat \alpha} T^{\hat a \ \hat \delta}_{\ \hat \beta} = T^{\hat a \ \hat \delta}_{\ \hat \alpha} T^{\hat a \ \hat \gamma}_{\ \hat \beta}$. While it would be natural to impose additional three-term identities for both color and numerator factors in the pseudo-real case, the numerators listed in Table \ref{prealrep} still obey the same set of two-term identities as in the complex case. It should be noted that these identities are not necessary for the consistency of the theory from the double-copy, i.e. the numerator identities stemming from the gauge-group generator commutation relations are sufficient for ensuring that the double-copy amplitudes obey the relevant Ward identities.  However, it might be possible to find different amplitude presentations which apply specifically to hypermultiplets in  pseudo-real representations and obey additional three-term identities in place of the two-term identities. As a consequence of the numerator relations and permutation symmetry, all numerators can be obtained in terms of a single master box numerator. \\

Finally, we note that the amplitude presentation in Table \ref{prealrep} can be extended to the case of an arbitrary number $n_H$ of half-hypermultiplets by dressing the numerator factors with Kronecker deltas with indices running over the number of half-hypermultiplets. For example, the first numerator is modified as
\begin{equation}
n_{1a}  =  {s^2 \delta^{IJ} \delta^{KL} \over \langle 12 \rangle \langle 34 \rangle} \delta^4 \big( Q \big) \ . 
\end{equation}
Additionally, bubble numerators with a matter loop acquire an extra factor of $n_H$.
This procedure leaves the numerator relations corresponding to (\ref{color-rel-basic}) unaltered, but the extra two-term relations are lost in the generic case.\footnote{Double-copying the numerators for amplitudes with $n_H$ external half-hypermultiplets with the ones in Table \ref{prealrep}, one can reproduce the $(1 + n_V/2)$ dependence of the divergence of the amplitudes of $\cN=4$ supergravity with $n_V$ vector multiplets \cite{Carrasco:2012ca}. This is an easy check of the consistency of our construction (the number of supergravity vector multiplets is related to the number of half-hypermultiplets as $ n_V = n_H + 2$).  } \\

\section{One-loop supergravity amplitudes from the double copy \label{sec4}}
\renewcommand{\theequation}{4.\arabic{equation}}
\setcounter{equation}{0}

We now focus on four-dimensional supergravity superamplitudes with four vector multiplets\footnote{By abuse of notation, we will use $\cal M$ to denote both superamplitudes and component amplitudes.}
\begin{equation}
{\cal M}^{1-\text{loop}}_4 \big(  1 {\cal V}_{a} , 2 {\cal V}_{b}, 3  {\cal V}_{c}, 4  {\cal V}_{d} \big)  \ , \label{grav-super}
\end{equation}
where $a,b,c,d$ are global indices running over the number of supergravity vectors that are realized as the product of two spin-$1/2$ asymptotic states. Note that these vectors transform as $SO(D-d)$ spinors under the global symmetry. 
These amplitudes can be obtained taking the double-copy of the amplitudes from the previous section with a four-fermion amplitude 
in the non-supersymmetric theory discussed in Section \ref{sec2},
\begin{equation}
{\cal A}^{1-\text{loop}}_4 \big(  1 \lambda_{a} , 2 \lambda_b, 
3  \lambda_c, 4  \lambda_d \big) \ .  \label{ampN0}
\end{equation}
The choice of chiralities for the external fermions determines whether a given external leg in (\ref{grav-super}) is associated to a $\cN=2$ vector on-shell superfield or its conjugate. 
We now focus on divergent contributions to the amplitude
(\ref{ampN0}). Since the non-supersymmetric theory is renormalizable, there is no UV-divergent box integral, 
and all the amplitude's divergences are linked to the divergences of three- and two-point Green functions. 
In particular, we write the one-loop corrections to vertices and 
inverse propagators as 
\begin{eqnarray}
i (\Pi^{1-\text{loop}}_\phi)^{\mu\nu, \hat a \hat b}(k^2) &=& -i g^2 \delta^{ \hat a \hat b}(\eta^{\mu\nu} - \eta_d^{\mu\nu} )  k^2 \tilde \Pi^{1-\text{loop}}_\phi(k^2) \ , \no \\
i (\Pi_A^{1-\text{loop}})^{\mu\nu, \hat a \hat b}(k^2) &=&  - i g^2 \delta^{\hat a \hat b}(k^2 \eta_d^{\mu\nu} - k^\mu k^\nu  )\tilde \Pi^{1-\text{loop}}_A(k^2) \ , \no \\
i {\cal V}^{1-\text{loop}}_{A \overline{\lambda} \lambda}(k_1,k_2,k_3) &=& 
 g^2{\cal V}^{\text{tree}}_{A \overline{\lambda}\lambda}(k_1,k_2,k_3) \tilde {\cal V}^{1-\text{loop}}_{A \overline{\lambda}\lambda}(k_1,k_2,k_3) \no \ , \\
i {\cal V}^{1-\text{loop}}_{\phi \overline{\lambda} \lambda}(k_1,k_2,k_3) &=& 
 g^2 {\cal V}^{\text{tree}}_{\phi \overline{\lambda}\lambda}(k_1,k_2,k_3) \tilde {\cal V}^{1-\text{loop}}_{\phi \overline{\lambda}\lambda}(k_1,k_2,k_3) \ .
\end{eqnarray}
Here we have split the $D$-dimensional gluons into $d$-dimensional gluons and $d$-dimensional scalars according to the values of the spacetime indices $\mu, \nu$. This is done by introducing the $d$-dimensional metric $\eta_d$  defined as 
\begin{equation}
 \eta_d = \text{diag} \big( \underbrace{1, -1, \ldots , -1}_{d} , \underbrace{0 ,\ldots , 0}_{D-d} \big)  \ .
\end{equation}
$\eta_d$ is generated by the integral reduction identities collected in Appendix \ref{AppRed}, since the loop momentum is taken to be in $d=4-2\epsilon$ dimensions.

We further split the tree-level, four-point amplitude in contributions corresponding to the three channels, 
\begin{equation}
{\cal A}^{\text{tree}}_4 = {\cal A}^{\text{tree}}_{s,A} + {\cal A}^{\text{tree}}_{s,\phi} + {\cal A}^{\text{tree}}_{t,A} + {\cal A}^{\text{tree}}_{t,\phi} + {\cal A}^{\text{tree}}_{u,A} +  {\cal A}^{\text{tree}}_{u,\phi}   \ ,
\end{equation}
where the vector and scalar channels have been written separately.
Taking into account the color factors in Table \ref{prealrep}, the integrated numerator factors of the non-supersymmetric theory are related to the part of the one-loop corrections to vertices and propagators which is proportional 
to the indices of the adjoint and matter representations:  
\begin{eqnarray}
\int {\tilde n_{4a} - \tilde n_{4b} \over D_4} \! &=& \!  A^{\text{tree}}_{s,\phi} \ \tilde{\cal V}^{1-\text{loop}}_{\phi \lambda \bar \lambda} (-k_1-k_2,k_1,k_2) \Big|_{T(G)} + A^{\text{tree}}_{s,A} \ \tilde{\cal V}^{1-\text{loop}}_{A \lambda \bar \lambda} (-k_1-k_2,k_1,k_2) \Big|_{T(G)} \no\ , \\
{1 \over S_{10}}\int {\tilde n_{10a} \over D_{10}} &=& {1 \over 2} A^{\text{tree}}_{s,A} \ \tilde{\Pi}_A^{1-\text{loop}} (s) \Big|_{T(G)}  + {1 \over 2} A^{\text{tree}}_{s,\phi} \ \tilde{\Pi}_\phi^{1-\text{loop}} (s) \Big|_{T(G)}\no \ , \\
{1 \over S_{10}}\int {\tilde n_{10b} \over D_{10}} &=&  {1 \over 2} A^{\text{tree}}_{s,A} \ \tilde{\Pi}_A^{1-\text{loop}} (s) \Big|_{T(R)}\no + {1 \over 2} A^{\text{tree}}_{s,\phi} \ \tilde{\Pi}_\phi^{1-\text{loop}} (s) \Big|_{T(R)} \ , 
\end{eqnarray}
where analogous relations for the other diagrams can be obtained by permutation.
Using the numerators in Table \ref{prealrep}, we have the following expression for the UV-divergent part of the supergravity amplitude,
\begin{eqnarray}
\label{first_m}
{\cal M}^{\text{1-loop}}\Big|_{\text{div}} \!\!\! \!\!\! &= \!\!\!& - { 2  s \delta^4(Q) \over \langle 12 \rangle \langle 34 \rangle   } \Big( {\kappa \over 2} \Big)^4
\left\{ s A^{\text{tree}}_{s,\phi} \Big( \tilde{\cal V}^{1-\text{loop}}_{\phi \lambda \bar \lambda}\Big|_{T(G),\text{div}} \! \! 
- {1\over 2} \tilde{\Pi}^{1-\text{loop}}_{\phi}\Big|_{T(G),\text{div}} +  {1\over 4} \tilde{\Pi}^{1-\text{loop}}_{\phi}\Big|_{T(R),\text{div}} \Big) 
\no \right. \\
&& \left. + s A^{\text{tree}}_{s,A} \Big( \tilde{\cal V}^{1-\text{loop}}_{A \lambda \bar \lambda}\Big|_{T(G),\text{div}}  - {1\over 2} \tilde{\Pi}^{1-\text{loop}}_{A}\Big|_{T(G),\text{div}} +  {1\over 4} \tilde{\Pi}^{1-\text{loop}}_{A}\Big|_{T(R),\text{div}} \Big) 
  \right\} + \text{Perms}. \qquad \label{gravity-div-1}
\end{eqnarray}

\begin{figure}
	\begin{center}	
		\includegraphics[width=0.31\textwidth]{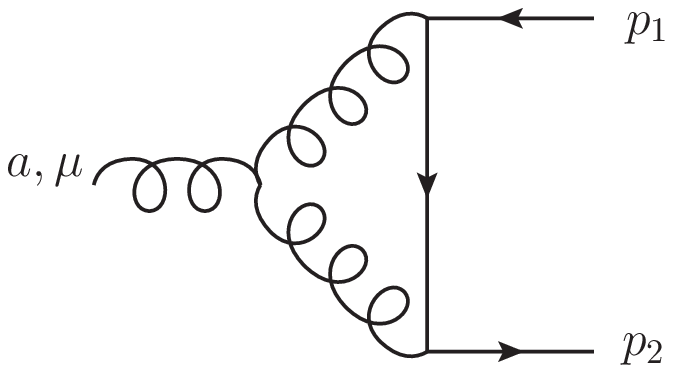} \hskip 2cm
		\includegraphics[width=0.31\textwidth]{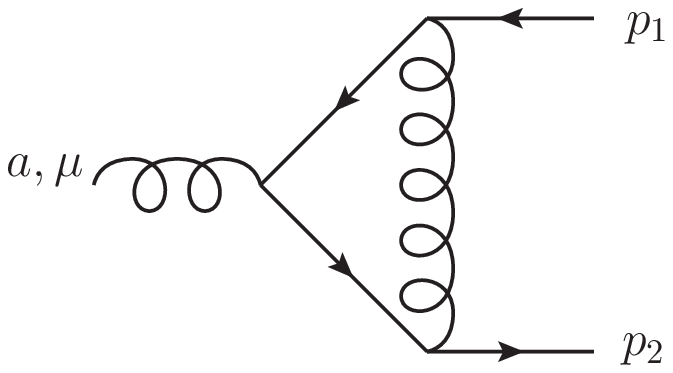}
		\caption{Triangle Feynman diagrams used in the computation for the non-supersymmetric gauge-theory factor. \label{fig-tri}}
	\end{center}
\end{figure}

We recognize that the combination of one-loop vertex and propagator corrections in (\ref{gravity-div-1}) corresponds to a piece of the beta function of the non-supersymmetric theory. The beta function $\beta(T(R),T(G),C(R))$ will formally depend on the index and quadratic Casimir of the representation, which are independent as long as no particular assumption is made on $R$. We can then rewrite the result above as
\begin{eqnarray}
{\cal M}^{\text{1-loop}}\Big|_{\text{div}} \!\! &=&  \! \! - {   s \delta^4(Q) \over \langle 12 \rangle \langle 34 \rangle   } \Big( {\kappa \over 2} \Big)^4 
\left\{ s A^{\text{tree}}_{s,\phi} \Big( \beta_\phi \big|_{T(G)} \! - \! { \beta_\phi \over 2} \Big|_{T(R)} \Big) 
\! + \! s A^{\text{tree}}_{s,A}  \Big( \beta_A \big|_{T(G)} \! - \! { \beta_A \over 2} \Big|_{T(R)} \Big) 
 \right\} {c_\Gamma \over \epsilon} \no \\ && \hskip 9cm  + \text{Perms},  
\end{eqnarray}
where $\beta_\phi \big|_{T(G)}$ denotes the part of the beta function for the Yukawa couplings which is proportional to $T(G)$. Contributions from the wave-function renormalization of the external fermions are proportional to $C(R)$ and hence do not appear in the above expression.

At one loop, the corrections to the vertices for the right-hand YM theory are 
obtained from the Feynman diagrams in Figure \ref{fig-tri} (see Appendix \ref{AppFeynman} for the Feynman rules employed in the calculation),
\begin{eqnarray}
i {\cal V}^{1-\text{loop}} \! \! \! &=& \! \! \!   g^3 (V T^bT^aT^b)  \frac{ \bar{u}_2 \Gamma^\nu (\slashed{l}+\slashed{p_1})\Gamma^\mu(\slashed{l}-\slashed{p_2})\Gamma_\nu u_1}{l^2(l+p_1)^2(l-p_2)^2} -
 \no \\ 
&& \! \! \!\! \! \!\! \! \!\! \! \!\! \! \!  i g^3 f^{abc} (V T^b T^c) \!  \big(\bar{u}_2 \Gamma_\nu\slashed{l}\Gamma_\rho u_1 \big)\! {(l+2p_1+p_2)^\rho\eta^{\mu\nu} \! + \!
	(l-p_1-2p_2)^\nu\eta^{\mu\rho} \!  + \! (p_2-2l-p_1)^\mu\eta^{\rho\nu} \over l^2(l+p_1)^2(l-p_2)^2} \no  . \\ \end{eqnarray}
This expression can be further simplified by employing the integral reduction identities collected in Appendix \ref{AppRed}. We obtain:
\begin{eqnarray}
\tilde {\cal V}^{1-\text{loop}}_{A \overline{\lambda}\lambda}(p_1,p_2) 
&=& 
-\frac{i}{2} C(R) \big\{4 I_2(p_1) + 4 I_2(p_2) + (D-10) I_2(p_1 + p_2) \big\} - i T(G) I_2(p_1 + p_2), \no\\
\tilde {\cal V}^{1-\text{loop}}_{\phi \overline{\lambda}\lambda}(p_1,p_2)  
&=&
-i C(R) \big\{2 I_2(p_1) + 2 I_2(p_2) + (D-6) I_2(p_1 + p_2)\big\} + \no \\ \  &&
\frac{i}{2} T(G) (D-4)   I_2(p_1 + p_2)  \ ,
\end{eqnarray}
where we have omitted a finite part proportional to the triangle integral $I_3(p_1,p_2)$. $I_2(k)$ denote bubble integrals. Bubble-on-external-leg integrals $I_2(p_1)$ and $I_2(p_2)$ have dropped out of the part of the vertex corrections which is proportional to $T(G)$. This implies that the  $1/\epsilon$ divergence of the vertex corrections that we will use in  (\ref{gravity-div-1}) is interpreted as a genuine UV divergence without any infrared contamination.

\begin{figure}
	\includegraphics[width=0.33\textwidth]{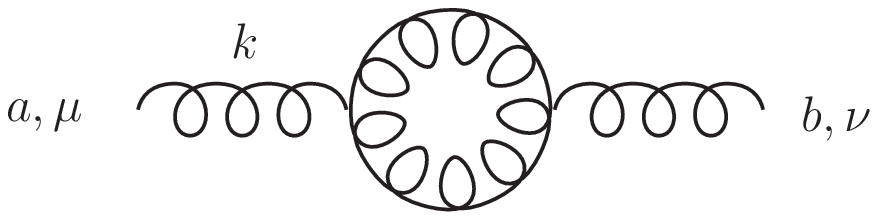}  \!\!\!\!\!
	\includegraphics[width=0.33\textwidth]{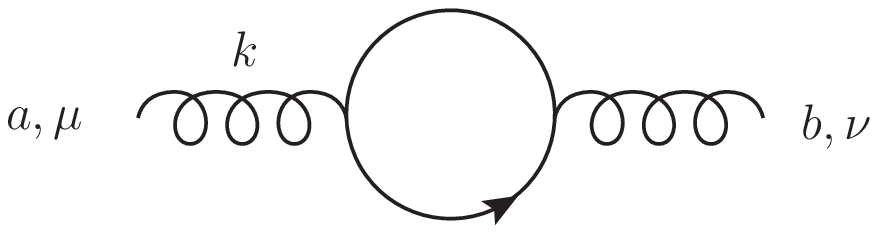}  \!\!\!\!\!
	\includegraphics[width=0.33\textwidth]{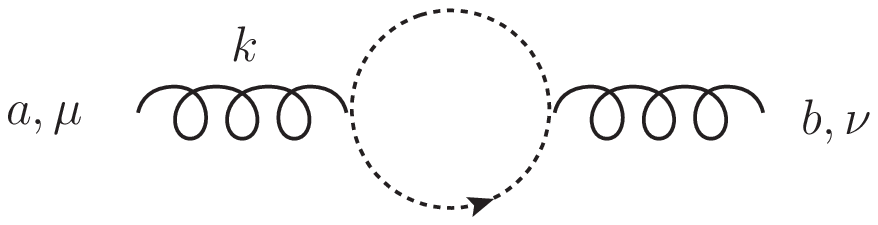}
	\caption{Bubble Feynman diagrams used in the computation for the non-supersymmetric gauge-theory factor. \label{fig-bub}}
\end{figure}

The calculation for the propagator corrections is performed along similar lines. The final result is:
\begin{eqnarray}
 \tilde \Pi^{1-\text{loop}}_\phi(k^2)
 &=&
i \big(2T(G)- n_F T(R) \big) I_2(k) \ , \no \\
\tilde \Pi^{1-\text{loop}}_A(k^2) 
&=&
- \frac{i}{6}\big\{(D-14)T(G)+4n_F T(R) \big\} I_2(k) \ , 
\end{eqnarray}
where $n_F$ is the number of four-dimensional fermions.
Equation (\ref{first_m}) then becomes
\begin{equation}
{\cal M}^{\text{1-loop}}\Big|_{\text{div}} \!= 
\!  { 2 i s \delta^4(Q) \over \langle 12 \rangle \langle 34 \rangle   } 
 \Big( {\kappa \over 2}\Big)^4 
\left\{ s A^{\text{tree}}_{s,\phi} 
\Big( 3- {D \over 2} + {n_F \over 4} \Big) 
+  s A^{\text{tree}}_{s,A} \Big( 
 {13 \over 6} - {D \over 12} +{n_F \over 6}
\Big) 
   \right\} {c_\Gamma \over \epsilon}  + \text{Perms} , \\
\end{equation}
where $c_\Gamma = i/(4 \pi)^2$. 
By setting $D=4+n_S$, where $n_S$ is the number of four-dimensional scalars, we obtain our master formula which expresses the value of the superamplitude's one-loop divergence in terms of the parameters of the construction for homogeneous supergravities:  
\begin{equation}
{\cal M}^{\text{1-loop}}\Big|_{\text{div}} \!\!\! = 
\! { 2 i s \delta^4(Q) \over \langle 12 \rangle \langle 34 \rangle   } \Big( {\kappa \over 2}\Big)^4 
\left\{ s A^{\text{tree}}_{s,\phi} 
\Big( 1 + {n_F \over 4} -  {n_S \over 2} \Big) 
+ 
s A^{\text{tree}}_{s,A} \Big( 
{11 \over 6} + {n_F \over 6 }- {n_S \over 12}
\Big) 
  \right\} {c_\Gamma\over\epsilon} + \text{Perms} . \label{master}  \\
\end{equation}

\subsection{Examples}

To simplify the expression (\ref{master}) further we specialize on amplitudes between four supergravity vectors,
\begin{equation}
{\cal M}^{\text{1-loop}}(1 A_{a-}, 2 A_{b-},3 A^c_+, 4 A^d_+) \! = \! {\cal A}^{\text{1-loop}}_{\cN=2}(1\chi_-, 2\chi_-,3 \chi_+, 4\chi_+) \otimes {\cal A}^{\text{1-loop}}_{\cN=0}(1\lambda_{a-}, 2\lambda_{b-},3 \lambda^c_{+}, 4\lambda^d_{+}),
\end{equation}
where the global indices of the last two vectors are raised with the inverse charge-conjugation matrix.
With this assignment of external polarizations and after using spinor-helicity identities which are collected in Appendix \ref{appConv}, we get the following expressions for tree amplitudes in the three channels,
\begin{eqnarray}
s A^\text{tree}_{s,\phi} &=& {i \over 2} \langle 12 \rangle [34] (\tilde C\tilde \Gamma^I)_{ab} (  \tilde \Gamma_I \tilde C^{-1})^{cd} \ , \\ 
t A^\text{tree}_{t,A} &=& i\langle 12 \rangle [34] \tilde \delta_{a}^d \delta_b^c \ , \\ 
u A^\text{tree}_{u,A} &=& i \langle 12 \rangle [34] \tilde \delta_{a}^c \delta_b^d  \ , 
\end{eqnarray}
where we have written the higher-dimensional Dirac and charge-conjugation matrices as $\Gamma^I = \gamma_5 \otimes \tilde \Gamma^I$ and $C={\cal C}_4 \otimes \tilde C$. In this case, the expression (\ref{master}) can be simplified as 
\begin{eqnarray}
{\cal M}^{\text{1-loop}}\Big|_{\text{div}} (1 A_{a-},2 A_{-b},3 A_{+}^c,4A_{+}^d) \!\!\! &=& 
\! { 2  \langle 12 \rangle^2 [ 34]^2 } \Big( {\kappa \over 2}\Big)^4 \no \\
&& \hskip -4.5cm \left\{ {1 \over 2} \Big( 1 + {n_F \over 4} -  {n_S \over 2} \Big) 
(\tilde C \tilde \Gamma^I)_{ab} ( \tilde \Gamma_I \tilde C^{-1})^{cd}
+ 
\Big( 
{11 \over 6} + {n_F \over 6 }- {n_S \over 12}
\Big) (\delta_{a}^d \delta_b^c + \delta_{a}^c \delta_b^d)
\right\} {c_\Gamma\over\epsilon} .  \no \\
\end{eqnarray}

We now consider some interesting particular cases. $n_S=0$ ($D=4$) corresponds to the so-called $\mathbb{CP}(n)$ or Luciani model \cite{Luciani:1977hp}. Supergravities in this family do not uplift to dimension higher than four and have symmetric scalar manifold
\begin{equation}
{ M}_{4D} = {U(1,n) \over U(1) \times U(n)} \ ,
\end{equation}
where $n$ is the number of vector multiplets.  The corresponding matter amplitudes between four vectors have one-loop divergence
\begin{equation}
{\cal M}^{\text{1-loop}}\Big|_{\text{div}} (1 A_{a-},2 A_{-b},3 A_{+}^c,4A_{+}^d) \!\!\! = 
\! { 2  \langle 12 \rangle^2 [ 34]^2 } \Big( {\kappa \over 2}\Big)^4  \left\{ 
\Big( 
{5 \over 3} + {n \over 6 }
\Big) (\delta_{a}^d \delta_b^c + \delta_{a}^c \delta_b^d)
\right\} {c_\Gamma\over\epsilon}   \ . \label{vectordiv}
\end{equation}
Another important example is the so-called Generic Jordan Family \cite{Gunaydin:1984ak, Gunaydin:1984ak}, 
which can be obtained by setting $D=6$ and keeping the number of 4D fermions arbitrary (with the relation $n=n_F+3$).\footnote{We recall that the Generic Jordan Family has two distinct double-copy realizations. The other realization involves a non-supersymmetric gauge theory with no matter fermions and an arbitrary number of scalars.} This theory has symmetric target space 
\begin{equation}
{ M}_{4D} = { SO(2,n-1) \over SO(n-1) \times SO(2)} \times {SU(1,1) \over U(1)} \ .
\end{equation}
Focusing on  amplitudes between two identical vectors and their CPT-conjugate states, we get the expression
\begin{equation}
{\cal M}^{\text{1-loop}}\Big|_{\text{div}} \!\!\! = 
\! { 2  \langle 12 \rangle^2 [ 34]^2 } \Big( {\kappa \over 2}\Big)^4  \left\{ 
\Big( 
{7 \over 3} + {n \over 3 }
\Big)
\right\} {c_\Gamma\over\epsilon}   \ .
\end{equation}
which reproduces the earlier result in ref. \cite{Carrasco:2012ca}.

By inspecting the two terms contributing to (\ref{master}), we see that the contribution corresponding to the vector exchange never vanishes.\footnote{One can also verify that the two terms cannot combine to give a vanishing divergence by contracting (\ref{vectordiv}) with $\delta^a_d \delta^b_c$. The resulting expression never vanishes for the values of the parameters given in Table \ref{tab1}. }  However, the contribution linked to the scalar exchange vanishes for 
\begin{eqnarray}
 D= 7, \ P=1 &\quad& (n_S= 3, \ n_F= 2) \ , \no \\ 
D= 8, \ P=1 &\quad& (n_S= 4, \ n_F= 4) \ , \no \\
D= 10, \ P=1 &\quad& (n_S= 6, \ n_F= 8) \ , \no \\
D= 14, \ P=1 &\quad& (n_S= 10, \ n_F= 16)  \ .
\end{eqnarray}
These are precisely the four Magical supergravities \cite{Gunaydin:1983rk,Gunaydin:1983bi}. Additionally, the contribution from the remaining channel matches the earlier computation in ref. \cite{Chiodaroli:2015wal}.\\

\section{Discussion}
\renewcommand{\theequation}{5.\arabic{equation}}
\setcounter{equation}{0}

In this paper, we have calculated the one-loop divergence for selected amplitudes between four vector multiplets in $\cN=2$ homogeneous Maxwell-Einstein supergravities with the double-copy construction, focusing  on amplitudes between vector constructed as the double copy of two spin-$1/2$ fields. Supergravity amplitudes are constructed using as building blocks gauge-theory amplitudes between four hypermultiplets in a presentation that obeys color/kinematics duality. Such amplitudes were first obtained in ref. \cite{Chiodaroli:2013upa} in terms of kinematical numerators which do not possess any explicit dependence on the loop momentum. Because of this property, the supergravity divergence is directly linked to the beta function of the non-supersymmetric gauge theory. In a sense, our calculation presents analogies with the one in ref. \cite{Bern:2012gh}, where the absence of some one- and two-loops divergences in half-maximal supergravity was linked to the renormalizability of the non-supersymmetric gauge theory entering the construction thanks to the absence of loop-momenta dependence in  $\cN=4$ sYM numerators at one and two loops.

Among the homogeneous supergravities, we do not find any matter amplitude which remains finite at one loop. An open question is how robust is this finding with respect to modifications of the construction. For example, we can generalize the construction by including $n_\phi$ complex matter scalars in the non-supersymmetric gauge theory. If the scalars are in the same pseudo-real representation as the half-hypermultiplets, the resulting supergravity theory will contain hypermultiplets in addition to the vector multiplets which are already present in the basic construction. 
The contribution to the supergravity divergence is modified by hypermultiplet loops and eq. (\ref{master}) becomes
\begin{equation}
{\cal M}\Big|_{\text{div}} \!\!\! = 
\! { 2 i s \delta^4(Q) \over \langle 12 \rangle \langle 34 \rangle   } \Big( {\kappa \over 2}\Big)^4 
\left\{ s A^{\text{tree}}_{s,\phi} 
\Big( 1 + {n_F \over 4} -  {n_S \over 2} \Big) 
+ 
s A^{\text{tree}}_{s,A} \Big( 
{11 \over 6} + {n_F \over 6 }- {n_S \over 12} + { n_\phi \over 12}
\Big) 
\right\} {c_\Gamma\over\epsilon} + \text{Perms} .   \\
\end{equation}
Hence, the additional contribution increases the divergence with respect to the Maxwell-Einstein case. Another possible modification is to add adjoint fermions. Since adjoint and matter representation contributions to (\ref{gravity-div-1}) come with opposite sign, adjoint fermions alleviate the UV divergence. However, they can be introduced in a way that is consistent with color/kinematics duality only if the gauge theory becomes supersymmetric \cite{Chiodaroli:2013upa}.

Our results can be understood in terms of counterterm analysis \cite{Elvang:2010jv,Bossard:2010bd, Bossard:2011tq, Beisert:2010jx, Elvang:2010xn, Bossard:2012xs, Freedman:2011uc, Bossard:2013rza, Brodel:2009hu, Kallosh:2011dp, Kallosh:2011qt, Kallosh:2012ei,Kallosh:2012yy}. The observed divergences correspond to the appearance of the linearized counterterms
\begin{equation}
 {\cal O}_1 =  \tilde C_{ac} \tilde C_{bd} (F^a_{\alpha \beta} F^{b\alpha \beta})( F^c_{\dot \alpha \dot \beta} F^{ d \dot \alpha \dot \beta}) \ , \qquad 
 {\cal O}_2 = \big( ( \tilde C \tilde \Gamma^I)_{ab} F^a_{\alpha \beta} F^{b\alpha \beta} \big)  \big( ( \tilde C \tilde \Gamma^I)_{cd} F^c_{\dot \alpha \dot \beta} F^{ d \dot \alpha \dot \beta}\big)  \ , \qquad 
\end{equation}
together with their supersymmetric completions. In the above equation, self-dual and anti-self-dual components of the vector field strengths in four dimensions are written using the two-component spinor notation as $F_{\alpha \beta} = {1 \over 2} F_{\mu \nu} \sigma^{\mu \nu}_{\alpha \beta}$. 
The second counterterm does not appear in the case of the Magical supergravities, which signals symmetry enhancement corresponding to the enlarged U-duality groups of these theories.

Finally, it would be interesting to see if the relation between supergravity divergences and physical quantities of the non-supersymmetric theory entering the construction (i.e. beta functions) can carry over to other matter amplitudes or higher loops. When supergravity vectors  constructed as vector times scalar are taken into account, the numerators for the supersymmetric gauge theory contain explicit dependence on the loop momentum, which makes it difficult to observe a relation between integrated quantities. As for extending the computation to higher loops, amplitudes which manifestly satisfy color/kinematics duality at two loops have recently become available \cite{Kalin:2018thp} and are likely to trigger further progress. \\

\section*{Acknowledgements}

M.C. would like to thank Murat G\"{u}naydin, Radu Roiban and Henrik Johansson for many insightful conversations and collaboration on related topics. We  thank Henrik Johansson and Murat G\"{u}naydin for valuable comments on an earlier draft of this paper. We also thank Gustav Mogull and Gregor K\"{a}lin for sharing a draft of their paper.
The research of M.C. is supported in part by the Knut and Alice Wallenberg Foundation under grant KAW 2013.0235 and the Swedish Research Council under grant 621-2014-5722. \\

   \newpage

\appendix

\section{Conventions \label{appConv}}

\renewcommand{\theequation}{A.\arabic{equation}}
\setcounter{equation}{0}

In this appendix we collect the conventions employed throughout this paper. Our notation can be obtained from the one of Elvang and Huang \cite{ElvangHuang} by replacing $\eta_{\mu\nu} \rightarrow - \eta_{\mu\nu}$. Our metric has mostly-minus signature and the Clifford algebra relation is
\begin{equation}
\{ \Gamma^\mu , \Gamma^\nu \} = 2 \eta^{\mu  \nu}  \ .
\end{equation}
$\Gamma^0$ is hermitian while the other gamma matrices are antihermitian. The four-dimensional gamma matrices $\gamma^\mu$ are
\begin{eqnarray}
\gamma^0 &= &  \sigma^1 \otimes 1 \ , \no \\
\gamma^1 &= & i \sigma^2 \otimes \sigma^1  \ , \no \\
\gamma^2 &= & i  \sigma^2 \otimes \sigma^2  \ , \no \\
\gamma^3 &= & i \sigma^2 \otimes \sigma^3 \ , \no \\
\gamma_5 &= &  \sigma^3 \otimes 1 \ .
\end{eqnarray}
Four-dimensional charge-conjugation and $B$ matrix are taken to be 
\begin{equation}
{\cal C}_4 =  \left( \begin{array}{cc}   \epsilon^{ \alpha \beta} & 0 \\ 0 & t_1 \epsilon_{\dot \alpha \dot \beta}  \end{array} \right) \ . \ , \qquad
B=  \left( \begin{array}{cc} 0 & t_1 \epsilon_{\dot \alpha \dot \beta} \\ \epsilon^{\a \b} & 0 \end{array} \right) \ .
\end{equation}
with $\epsilon^{12}= - \epsilon_{12} = +1$. $t_1$ is a sign to be assigned according to the value of the parameter $D$ in the construction. The charge conjugation matrix obeys the conditions ${\cal C}_4^t = - {\cal C}_4$ and $(\gamma^\mu)^t = -t_1 {\cal C}_4^{-1} \gamma^\mu {\cal C}_4$.

Higher dimensional gamma matrices are written as
\begin{eqnarray}
\Gamma^\mu &=& \gamma^\mu \otimes {\bf 1} \ , \qquad \mu < 4 \ , \no \\
\Gamma^\mu &=& \gamma_5 \otimes \tilde \Gamma^{\mu} \ , \qquad \mu \geq 4 \ . 
\end{eqnarray}
We introduce indices $I,J$ running over the internal dimensions.
$\tilde C$ and $\tilde \Gamma^I$ denote the components of the charge-conjugation matrix acting on the spinor indices corresponding to the internal $(D-d)$ dimensions. The sign $t_1$ is fixed by the requirement that $\tilde C \tilde \Gamma^I$ is always symmetric or, alternatively, that $C \Gamma^\mu$ is always antisymmetric.  

Introducing the spinor-helicity variables as
\begin{equation}
\lambda(p) = \left( \begin{array}{c}
|p ]_\a \\ |p\rangle^{\dot \a}
\end{array} \right) \ ,
\end{equation}
the Majorana condition is rewritten as
\begin{equation}
\lambda^* = B \lambda  \quad \rightarrow \quad \left\{ \begin{array}{c}
\big( |p]_\a \big)^* = \epsilon_{\dot \a \dot \b} |p\rangle^{\dot \b} = |p\rangle_{\dot \a} \\
\big( |p \rangle^{\dot \a} \big)^* = \epsilon^{\a \b} |p]_{ \b} = |p]^{ \a} \\
\end{array} \right. \ .
\end{equation}
With this condition we have
\begin{equation}
([pq])^* = \langle qp \rangle
\end{equation}
for real momenta. We can  expand null momenta using spinor-helicity variables as
\begin{equation}
\cancel{p} = - |p\rangle [p| - |p]\langle p| \ .
\end{equation}
We also have the identities
\begin{eqnarray}
\langle pq \rangle [qp] &=& 2 p\cdot q \ ,\\
\langle p |\gamma^\mu| q ] \langle r |\gamma_\mu| s ] &=& 2 \langle pr \rangle [sq] \ .
\end{eqnarray}

\section{Feynman Rules \label{AppFeynman}}

\renewcommand{\theequation}{B.\arabic{equation}}
\setcounter{equation}{0}

In this appendix, we collect the Feynman rules for the non-supersymmetric gauge theory entering the double-copy construction, which are obtained from the Lagrangian (\ref{L-left}). All momenta are taken as in-going.
\begin{eqnarray} 
&&\text{Fermion propagator: } 
\begin{array}{l} \\
\includegraphics[scale=0.6]{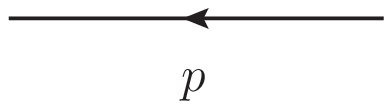}
\end{array}
=\frac{ip_\mu\Gamma^\mu }{p^2}C^{-1}V^{-1} \ , \\
&&\text{Gluon propagator:  }  
\begin{array}{l} \\
\includegraphics[scale=0.6]{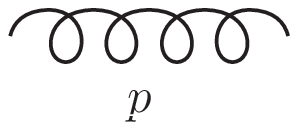}
\end{array}
=\frac{-i\eta_{\mu\nu}}{p^2} \ , \\
&&\text{Ghost propagator:  }
\begin{array}{l} \\
\includegraphics[scale=0.6]{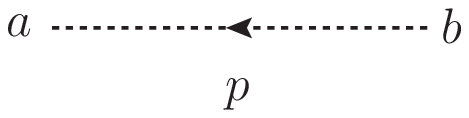}
\end{array}
=\frac{-i\delta^{ab}}{p^2} \ , \\
&&\text{Fermion vertex:  } 
\begin{array}{l} 
\includegraphics[scale=0.6]{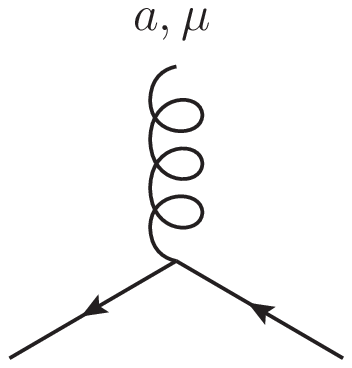}
\end{array}
=ig Vt^a C\Gamma^\mu \\
&&\text{Gluon vertex: }
\begin{array}{l} 
\includegraphics[scale=0.6]{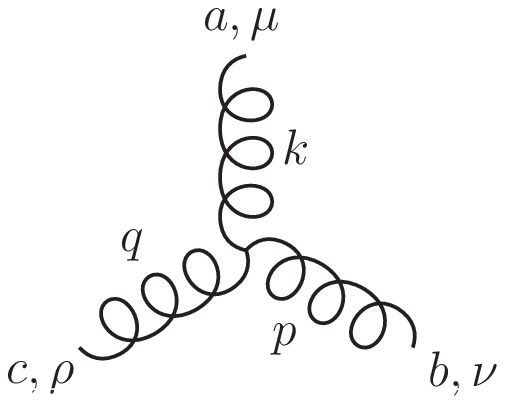}
\end{array} \! \! \! \! \! \!  
=gf^{abc}\big(
\eta^{\mu\nu}(k-p)^\rho
+\eta^{\nu\rho}(p-q)^\mu
+\eta^{\rho\mu}(q-k)^\nu
\big)  , \qquad \ \\
&&\text{Ghost vertex:  } 
\begin{array}{l} 
\includegraphics[scale=0.6]{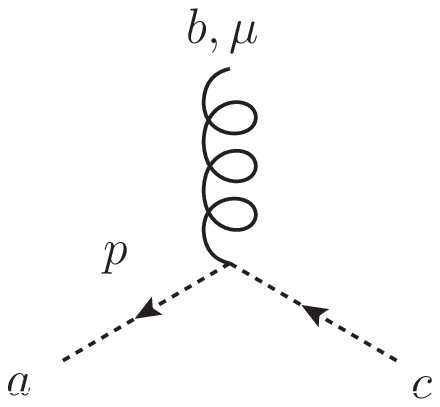}
\end{array}
=gf^{abc}p^\mu \ .
\end{eqnarray}

\section{Details on the orbifold numerators \label{Apporbifold}}

\renewcommand{\theequation}{C.\arabic{equation}}
\setcounter{equation}{0}

To obtain an explicit presentation of the one-loop (super)amplitude with four external hypermultiplets in a complex representation,
it is convenient to start from the amplitude specified in eq. (5.46) of ref. \cite{Chiodaroli:2013upa} and perform the summation 
over the orbifold group elements. We consider a $\mathbb{Z}_3$ orbifold taking $SU(3N)$ as the gauge group for the parent theory.
The orbifold action is given by
\begin{equation}
r_n = \left( \begin{array}{cccc} 1 & 0 & 0 & 0 \\   
0 & 1 & 0& 0 \\
0 & 0 & e^{i n \theta} & 0 \\
0 & 0 & 0 & e^{-i n \theta}   \end{array}
\right) \ , \qquad   g_n = \left( \begin{array}{ccc} I_N & 0 & 0  \\   
0 & e^{i n \theta} I_N & 0 \\
0 & 0 & e^{2 i n \theta} I_N   \end{array}
\right) \ , \qquad \theta = { 2 \pi \over 3 }   \ .
\end{equation}
This choice breaks the gauge group as $SU(3N) \rightarrow SU(N)^3 \times U(1)^2$.
We then split accordingly the $SU(3N)$ adjoint indices as $\hat A=( \hat a, \hat \alpha, \hat{\bar \alpha})$, 
where the indices run over the following representations of the (product) gauge group,
\begin{eqnarray}
\hat a: & \qquad& ({\bf N}^2 -1 ,1 ,1) \oplus (1,{\bf N}^2 -1  ,1) \oplus (1,1,{\bf N}^2 -1) \oplus 2 (1 ,1 ,1) \ , \no \\
{ \bar \alpha}: & \qquad& ({\bf N} , \bar {\bf N}  ,1) \oplus (1 ,{\bf N}  ,\bar {\bf N}) \oplus (\bar {\bf N},1, {\bf N} ) \ , \no \\
\hat{\bar \alpha}: & \qquad& (\bar {\bf N} , {\bf N}  ,1) \oplus (1 ,\bar {\bf N}  ,{\bf N}) \oplus ({\bf N},1,\bar {\bf N} ) 
\ .  
\end{eqnarray}
Projectors into the three sets of representations are written as
\begin{eqnarray}
({\cal P}_G \Phi)^{\hat A} &=& \sum_{\Gamma} g^{\hat A \hat B} \Phi^{\hat B} = \big( \Phi^{\hat a}, 0 , 0 \big) \no \ , \\ 
({\cal P}_R \Phi)^{\hat A} &=& \sum_{\Gamma} r^3_3 g^{\hat A \hat B} \Phi^{\hat B} = \big(0, \Phi_{\hat \alpha} , 0 \big) \ , \no \\
({\cal P}_{\overline{R}} \Phi)^{\hat A} &=& \sum_{\Gamma} r^4_4 g^{\hat A \hat B} \Phi^{\hat B} = \big(0,0, \Phi^{\hat \alpha} \big) \ ,
\end{eqnarray}
where $\Phi$ is a generic field of the parent theory. 
Representation matrices of the $R$ representation are then given by
\begin{equation}
\tilde T^{\hat a \ \hat \beta}_{\ \hat \alpha} = - ({\cal P}_R)_{\hat \alpha}^{\ \hat A} ({\cal P}_{\overline{R}})^{\hat \beta \hat B} 
({\cal P}_G)^{\hat a \hat C} \tilde f^{\hat A \hat B \hat C} \ .
\end{equation}
At four points, supersymmetry implies that amplitudes with four external hypermultiplet fields can be organized in superamplitudes which can be directly obtained from the amplitudes between two identical scalars and their conjugates given in ref. \cite{Chiodaroli:2013upa},
\begin{equation}
{\cal F}_4^{1-\text{loop}} \big( 1 {\cal Q} , 2 {\cal Q}, 3 \overline{ \cal Q}, 4 \overline{\cal Q } \big) = {\delta^4 \big( \sum_i \eta_i^\alpha |i\rangle \big) \over \langle 12 \rangle \langle 34 \rangle}
{\cal A}_4^{1-\text{loop}} \big( 1 \varphi , 2 \varphi, 3 \bar \varphi, 4 \bar \varphi \big) \ .
\end{equation}
Component amplitudes can be easily extracted my acting with derivatives with respect to the Grassmann variables $\eta$.

It should be noted that the amplitudes in \cite{Chiodaroli:2013upa} were obtained with a procedure that is not sensitive to bubble-on-external-leg graphs. In principle, it is possible to add back these graphs  in a way that preserves color/kinematics duality by adding to all numerators terms proportional to the squares of the external momenta $p_i^2$. When the external momenta are put on-shell, i.e. the limit $p_i^2\rightarrow 0$ is taken, the additional contributions drop out of the final expression in all graphs except the ones with bubbles on one external leg, which have a $1/p_i^2$ factor in the propagators. However, these graphs can be safely ignored in the present context as they do not contribute to the gravity amplitudes because each of the two numerators entering the double-copy formula is proportional to $p_i^2$. At the level of gauge-theory amplitudes, bubbles-on-external-legs integrals vanish in dimensional regularization. However, they can lead to non-vanishing contributions if particular kinematical limits (UV or infrared) are inspected.

\section{Integral Reduction \label{AppRed}}

\renewcommand{\theequation}{D.\arabic{equation}}
\setcounter{equation}{0}

\begin{figure}[t]
	\begin{center}	
		\includegraphics[scale=0.7]{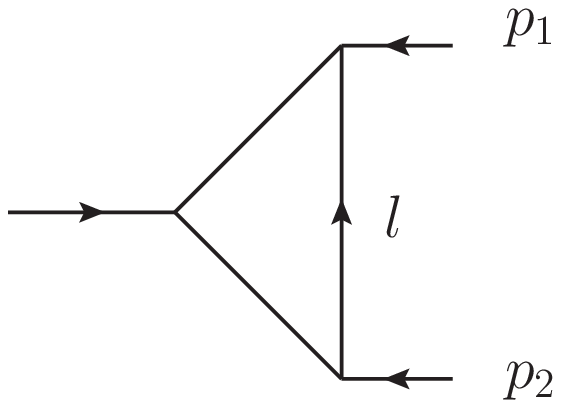} \hskip 2cm
		\includegraphics[scale=0.7]{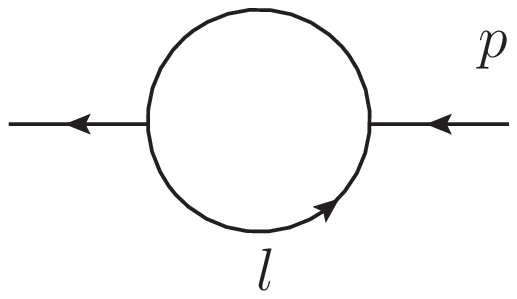}
		\caption{Orientation for external and loop momenta used in the integral reduction identities.}
		\label{fig-orientation}
	\end{center}
\end{figure}
Orientation of loop and external  momenta are taken as shown in Figure \ref{fig-orientation}. The integral-reduction identities for bubble diagrams are:
\begin{eqnarray}
 I_2(l^\mu;p)&=& -\frac{1}{2}I_2(p)p^\mu  \ , \\ 
I_2(l^\mu l^\nu;p) &=& 
\frac{d}{4d-4}I_2(p)p^\mu p^\nu 
-\frac{p^2}{4d-4}I_2(p)\eta_d^{\mu\nu} \ .
\end{eqnarray}
For triangle diagrams, integral reduction identities are as follows:
\begin{eqnarray}
	I_3(l^\mu;p_1,p_2) &=&
	\frac{p_1^\mu (I_2(p_1+p_2)-I_2(p_1))}{2p_1\cdot p_2}
	+\frac{p_2^\mu(I_2(1,p_2)-I_2(p_1+p_2) )}{2p_1\cdot p_2} \ , \\
	I_3(l^\mu l^\nu;p_1,p_2) &=&
	\frac{\eta_d^{\mu\nu}I_2(p_1+p_2)}{2(d-2)}
	+
	p_1^\mu p_1^\nu \frac{I_2(p_1) -I_2(p_1+p_2)}{4p_1\cdot p_2} 
	+ \no \\
	&& p_2^\mu p_2^\nu \frac{I_2(p_2)-I_2(p_1+p_2)}{4p_1\cdot p_2}
	+\no \\
	&&p_1^{(\mu}p_2^{\nu)}\frac{(d-4)I_2(p_1+p_2)}{4(d-2)p_1\cdot p_2}
\end{eqnarray}
Note that the above identities depend on the $d$-dimensional metric $\eta_d^{\mu  \nu}$.


\begin{thebibliography}{100}

\bibitem{Bern:2008qj} Z.~Bern, J.~J.~M.~Carrasco and H.~Johansson, 
``New Relations for Gauge-Theory Amplitudes,'' 
Phys.\ Rev.\ D {\bf 78}, 085011 (2008) 
[arXiv:0805.3993 [hep-ph]]. 


\bibitem{Bern:2010ue} Z.~Bern, J.~J.~M.~Carrasco and H.~Johansson, 
``Perturbative Quantum Gravity as a Double Copy of Gauge Theory,'' 
Phys.\ Rev.\ Lett.\ {\bf 105}, 061602 (2010) 
[arXiv:1004.0476 [hep-th]]. 

\bibitem{Bern:2017yxu} Z.~Bern, J.~J.~Carrasco, W.~M.~Chen, H.~Johansson and R.~Roiban,
``Gravity Amplitudes as Generalized Double Copies of Gauge-Theory Amplitudes,''
Phys.\ Rev.\ Lett.\  {\bf 118}, no. 18, 181602 (2017)
doi:10.1103/PhysRevLett.118.181602
[arXiv:1701.02519 [hep-th]].

\bibitem{Bern:2017ucb} Z.~Bern, J.~J.~M.~Carrasco, W.~M.~Chen, H.~Johansson, R.~Roiban and M.~Zeng,
``Five-loop four-point integrand of $N=8$ supergravity as a generalized double copy,''
Phys.\ Rev.\ D {\bf 96}, no. 12, 126012 (2017)
doi:10.1103/PhysRevD.96.126012
[arXiv:1708.06807 [hep-th]].

\bibitem{Bern:2018jmv} Z.~Bern, J.~J.~Carrasco, W.~M.~Chen, A.~Edison, H.~Johansson, J.~Parra-Martinez, R.~Roiban and M.~Zeng,
``Ultraviolet Properties of $\mathcal N = 8$ Supergravity at Five Loops,''
Phys.\ Rev.\ D {\bf 98}, no. 8, 086021 (2018)
doi:10.1103/PhysRevD.98.086021
[arXiv:1804.09311 [hep-th]].

\bibitem{Bern:2012cd} Z.~Bern, S.~Davies, T.~Dennen and Y.~t.~Huang,
``Absence of Three-Loop Four-Point Divergences in N=4 Supergravity,''
Phys.\ Rev.\ Lett.\  {\bf 108}, 201301 (2012)
doi:10.1103/PhysRevLett.108.201301
[arXiv:1202.3423 [hep-th]].

\bibitem{Bern:2012gh} Z.~Bern, S.~Davies, T.~Dennen and Y.~t.~Huang,
``Ultraviolet Cancellations in Half-Maximal Supergravity as a Consequence of the Double-Copy Structure,''
Phys.\ Rev.\ D {\bf 86}, 105014 (2012)
doi:10.1103/PhysRevD.86.105014
[arXiv:1209.2472 [hep-th]].

\bibitem{Bern:2013uka} Z.~Bern, S.~Davies, T.~Dennen, A.~V.~Smirnov and V.~A.~Smirnov,
``Ultraviolet Properties of N=4 Supergravity at Four Loops,''
Phys.\ Rev.\ Lett.\  {\bf 111}, no. 23, 231302 (2013)
doi:10.1103/PhysRevLett.111.231302
[arXiv:1309.2498 [hep-th]].

\bibitem{Bern:2014sna} Z.~Bern, S.~Davies and T.~Dennen,
``Enhanced ultraviolet cancellations in $\mathcal N=5$ supergravity at four loops,''
Phys.\ Rev.\ D {\bf 90}, no. 10, 105011 (2014)
doi:10.1103/PhysRevD.90.105011
[arXiv:1409.3089 [hep-th]].

\bibitem{Bern:2017lpv} Z.~Bern, M.~Enciso, J.~Parra-Martinez and M.~Zeng,
``Manifesting enhanced cancellations in supergravity: integrands versus integrals,''
JHEP {\bf 1705}, 137 (2017)
doi:10.1007/JHEP05(2017)137
[arXiv:1703.08927 [hep-th]].

\bibitem{Carrasco:2013ypa} J.~J.~M.~Carrasco, R.~Kallosh, R.~Roiban and A.~A.~Tseytlin,
``On the U(1) duality anomaly and the S-matrix of N=4 supergravity,''
JHEP {\bf 1307}, 029 (2013)
doi:10.1007/JHEP07(2013)029
[arXiv:1303.6219 [hep-th]].

\bibitem{Bern:2017rjw} Z.~Bern, J.~Parra-Martinez and R.~Roiban,
``Canceling the U(1) Anomaly in the $S$ Matrix of $N$=4 Supergravity,''
Phys.\ Rev.\ Lett.\  {\bf 121}, no. 10, 101604 (2018)
doi:10.1103/PhysRevLett.121.101604
[arXiv:1712.03928 [hep-th]].

\bibitem{Bern:2011rj} Z.~Bern, C.~Boucher-Veronneau and H.~Johansson, 
``N $\ge$ 4 Supergravity Amplitudes from Gauge Theory at One Loop,'' 
Phys.\ Rev.\ D {\bf 84}, 105035 (2011) 
[arXiv:1107.1935 [hep-th]]. 

\bibitem{Johansson:2014zca} 
H.~Johansson and A.~Ochirov,
``Pure Gravities via Color-Kinematics Duality for Fundamental Matter,''
JHEP {\bf 1511}, 046 (2015)
doi:10.1007/JHEP11(2015)046
[arXiv:1407.4772 [hep-th]].

\bibitem{Carrasco:2012ca} J.~J.~M.~Carrasco, M.~Chiodaroli, M.~G{u}naydin and R.~Roiban, 
``One-loop four-point amplitudes in pure and matter-coupled N $\le$ 4 supergravity,'' 
JHEP {\bf 1303}, 056 (2013) 
[arXiv:1212.1146 [hep-th]]; 

\bibitem{Bern:2013yya} Z.~Bern, S.~Davies, T.~Dennen, Y.~t.~Huang and J.~Nohle,
``Color-Kinematics Duality for Pure Yang-Mills and Gravity at One and Two Loops,''
Phys.\ Rev.\ D {\bf 92}, no. 4, 045041 (2015)
[arXiv:1303.6605 [hep-th]].

\bibitem{Chiodaroli:2015wal} M.~Chiodaroli, M.~Gunaydin, H.~Johansson and R.~Roiban,
``Complete construction of magical, symmetric and homogeneous N=2 supergravities as double copies of gauge theories,''
arXiv:1512.09130 [hep-th].

\bibitem{Anastasiou:2016csv} A.~Anastasiou, L.~Borsten, M.~J.~Duff, M.~J.~Hughes, A.~Marrani, S.~Nagy and M.~Zoccali,
``Twin supergravities from Yang-Mills theory squared,''
Phys.\ Rev.\ D {\bf 96}, no. 2, 026013 (2017)
doi:10.1103/PhysRevD.96.026013
[arXiv:1610.07192 [hep-th]].

\bibitem{Anastasiou:2017nsz} A.~Anastasiou, L.~Borsten, M.~J.~Duff, A.~Marrani, S.~Nagy and M.~Zoccali,
``Are all supergravity theories Yang–Mills squared?,''
Nucl.\ Phys.\ B {\bf 934}, 606 (2018)
doi:10.1016/j.nuclphysb.2018.07.023
[arXiv:1707.03234 [hep-th]].

\bibitem{Johansson:2017bfl} H.~Johansson, G.~K\"{a}lin and G.~Mogull,
``Two-loop supersymmetric QCD and half-maximal supergravity amplitudes,''
JHEP {\bf 1709}, 019 (2017)
doi:10.1007/JHEP09(2017)019
[arXiv:1706.09381 [hep-th]].



\bibitem{Chiodaroli:2014xia} M.~Chiodaroli, M.~Gunaydin, H.~Johansson and R.~Roiban,
``Scattering amplitudes in $ \mathcal{N}=2 $ Maxwell-Einstein and Yang-Mills/Einstein supergravity,''
JHEP {\bf 1501}, 081 (2015)
doi:10.1007/JHEP01(2015)081
[arXiv:1408.0764 [hep-th]].

\bibitem{Chiodaroli:2015rdg} M.~Chiodaroli, M.~Gunaydin, H.~Johansson and R.~Roiban,
``Spontaneously Broken Yang-Mills-Einstein Supergravities as Double Copies,''
JHEP {\bf 1706}, 064 (2017)
doi:10.1007/JHEP06(2017)064
[arXiv:1511.01740 [hep-th]].

\bibitem{Chiodaroli:2016jqw} 
M.~Chiodaroli,
``Simplifying amplitudes in Maxwell-Einstein and Yang-Mills-Einstein supergravities,''
doi:10.1515/9783110452150-011
arXiv:1607.04129 [hep-th].

\bibitem{Chiodaroli:2017ehv} M.~Chiodaroli, M.~Gunaydin, H.~Johansson and R.~Roiban,
``Gauged Supergravities and Spontaneous Supersymmetry Breaking from the Double Copy Construction,''
Phys.\ Rev.\ Lett.\  {\bf 120}, no. 17, 171601 (2018)
doi:10.1103/PhysRevLett.120.171601
[arXiv:1710.08796 [hep-th]].

\bibitem{Johansson:2017srf} H.~Johansson and J.~Nohle,
``Conformal Gravity from Gauge Theory,''
arXiv:1707.02965 [hep-th].

\bibitem{Johansson:2018ues} 
H.~Johansson, G.~Mogull and F.~Teng,
``Unraveling conformal gravity amplitudes,''
JHEP {\bf 1809}, 080 (2018)
doi:10.1007/JHEP09(2018)080
[arXiv:1806.05124 [hep-th]].

\bibitem{Johansson:2015oia} H.~Johansson and A.~Ochirov,
``Color-Kinematics Duality for QCD Amplitudes,''
JHEP {\bf 1601}, 170 (2016)
doi:10.1007/JHEP01(2016)170
[arXiv:1507.00332 [hep-ph]].

\bibitem{delaCruz:2015dpa} L.~de la Cruz, A.~Kniss and S.~Weinzierl,
``Proof of the fundamental BCJ relations for QCD amplitudes,''
JHEP {\bf 1509}, 197 (2015)
doi:10.1007/JHEP09(2015)197
[arXiv:1508.01432 [hep-th]].

\bibitem{Chen:2013fya} G.~Chen and Y.~J.~Du,
``Amplitude Relations in Non-linear Sigma Model,''
JHEP {\bf 1401}, 061 (2014)
doi:10.1007/JHEP01(2014)061
[arXiv:1311.1133 [hep-th]].

\bibitem{Chen:2014dfa} G.~Chen, Y.~J.~Du, S.~Li and H.~Liu,
``Note on off-shell relations in nonlinear sigma model,''
JHEP {\bf 1503}, 156 (2015)
doi:10.1007/JHEP03(2015)156
[arXiv:1412.3722 [hep-th]].

\bibitem{Cheung:2016prv} C.~Cheung and C.~H.~Shen,
``Symmetry for Flavor-Kinematics Duality from an Action,''
Phys.\ Rev.\ Lett.\  {\bf 118}, no. 12, 121601 (2017)
doi:10.1103/PhysRevLett.118.121601
[arXiv:1612.00868 [hep-th]].

\bibitem{Du:2016tbc} Y.~J.~Du and C.~H.~Fu,
``Explicit BCJ numerators of nonlinear simga model,''
JHEP {\bf 1609}, 174 (2016)
doi:10.1007/JHEP09(2016)174
[arXiv:1606.05846 [hep-th]].

\bibitem{Chen:2016zwe} G.~Chen, S.~Li and H.~Liu,
``Off-shell BCJ Relation in Nonlinear Sigma Model,''
arXiv:1609.01832 [hep-th].

\bibitem{Chiodaroli:2017ngp} M.~Chiodaroli, M.~Gunaydin, H.~Johansson and R.~Roiban,
``Explicit Formulae for Yang-Mills-Einstein Amplitudes from the Double Copy,''
JHEP {\bf 1707}, 002 (2017)
doi:10.1007/JHEP07(2017)002
[arXiv:1703.00421 [hep-th]].

\bibitem{Stieberger:2016lng} S.~Stieberger and T.~R.~Taylor,
``New relations for Einstein–Yang–Mills amplitudes,''
Nucl.\ Phys.\ B {\bf 913}, 151 (2016)
doi:10.1016/j.nuclphysb.2016.09.014
[arXiv:1606.09616 [hep-th]].

\bibitem{Stieberger:2014cea} S.~Stieberger and T.~R.~Taylor,
``Graviton as a Pair of Collinear Gauge Bosons,''
Phys.\ Lett.\ B {\bf 739}, 457 (2014)
doi:10.1016/j.physletb.2014.10.057
[arXiv:1409.4771 [hep-th]].

\bibitem{Stieberger:2015qja} S.~Stieberger and T.~R.~Taylor,
``Graviton Amplitudes from Collinear Limits of Gauge Amplitudes,''
Phys.\ Lett.\ B {\bf 744}, 160 (2015)
doi:10.1016/j.physletb.2015.03.053
[arXiv:1502.00655 [hep-th]].

\bibitem{Nandan:2016pya} D.~Nandan, J.~Plefka, O.~Schlotterer and C.~Wen,
``Einstein-Yang-Mills from pure Yang-Mills amplitudes,''
JHEP {\bf 1610}, 070 (2016)
doi:10.1007/JHEP10(2016)070
[arXiv:1607.05701 [hep-th]].

\bibitem{delaCruz:2016gnm} L.~de la Cruz, A.~Kniss and S.~Weinzierl,
``Relations for Einstein–Yang–Mills amplitudes from the CHY representation,''
Phys.\ Lett.\ B {\bf 767}, 86 (2017)
doi:10.1016/j.physletb.2017.01.036
[arXiv:1607.06036 [hep-th]].

\bibitem{Teng:2017tbo} F.~Teng and B.~Feng,
``Expanding Einstein-Yang-Mills by Yang-Mills in CHY frame,''
JHEP {\bf 1705}, 075 (2017)
doi:10.1007/JHEP05(2017)075
[arXiv:1703.01269 [hep-th]].

\bibitem{Du:2017gnh} Y.~J.~Du, B.~Feng and F.~Teng,
``Expansion of All Multitrace Tree Level EYM Amplitudes,''
JHEP {\bf 1712}, 038 (2017)
doi:10.1007/JHEP12(2017)038
[arXiv:1708.04514 [hep-th]].

\bibitem{Nandan:2018ody} D.~Nandan, J.~Plefka and G.~Travaglini,
``All rational one-loop Einstein-Yang-Mills amplitudes at four points,''
JHEP {\bf 1809}, 011 (2018)
doi:10.1007/JHEP09(2018)011
[arXiv:1803.08497 [hep-th]].


\bibitem{Stieberger:2009hq} S.~Stieberger, 
``Open $\&$ Closed vs. Pure Open String Disk Amplitudes,'' 
arXiv:0907.2211 [hep-th]. 

\bibitem{BjerrumBohr:2009rd} N.~E.~J.~Bjerrum-Bohr, P.~H.~Damgaard and P.~Vanhove,
``Minimal Basis for Gauge Theory Amplitudes,''
Phys.\ Rev.\ Lett.\  {\bf 103}, 161602 (2009)
doi:10.1103/PhysRevLett.103.161602
[arXiv:0907.1425 [hep-th]].

\bibitem{BjerrumBohr:2010zs} N.~E.~J.~Bjerrum-Bohr, P.~H.~Damgaard, T.~Sondergaard and P.~Vanhove,
``Monodromy and Jacobi-like Relations for Color-Ordered Amplitudes,''
JHEP {\bf 1006}, 003 (2010)
doi:10.1007/JHEP06(2010)003
[arXiv:1003.2403 [hep-th]].

\bibitem{Tourkine:2016bak} P.~Tourkine and P.~Vanhove,
``Higher-loop amplitude monodromy relations in string and gauge theory,''
Phys.\ Rev.\ Lett.\  {\bf 117}, no. 21, 211601 (2016)
doi:10.1103/PhysRevLett.117.211601
[arXiv:1608.01665 [hep-th]].

\bibitem{Hohenegger:2017kqy} S.~Hohenegger and S.~Stieberger,
``Monodromy Relations in Higher-Loop String Amplitudes,''
Nucl.\ Phys.\ B {\bf 925}, 63 (2017)
doi:10.1016/j.nuclphysb.2017.09.020
[arXiv:1702.04963 [hep-th]].

\bibitem{Fu:2018hpu} C.~H.~Fu, P.~Vanhove and Y.~Wang,
``A Vertex Operator Algebra Construction of the Colour-Kinematics Dual numerator,''
JHEP {\bf 1809}, 141 (2018)
doi:10.1007/JHEP09(2018)141
[arXiv:1806.09584 [hep-th]].

\bibitem{Mafra:2011kj} C.~R.~Mafra, O.~Schlotterer and S.~Stieberger, 
``Explicit BCJ Numerators from Pure Spinors,'' 
JHEP {\bf 1107}, 092 (2011) 
[arXiv:1104.5224 [hep-th]].

\bibitem{Mafra:2014gja} C.~R.~Mafra and O.~Schlotterer,
``Towards one-loop SYM amplitudes from the pure spinor BRST cohomology,''
Fortsch.\ Phys.\  {\bf 63}, no. 2, 105 (2015)
doi:10.1002/prop.201400076
[arXiv:1410.0668 [hep-th]].

\bibitem{He:2015wgf} S.~He, R.~Monteiro and O.~Schlotterer,
``String-inspired BCJ numerators for one-loop MHV amplitudes,''
JHEP {\bf 1601}, 171 (2016)
[arXiv:1507.06288 [hep-th]].

\bibitem{Mafra:2015mja} C.~R.~Mafra and O.~Schlotterer,
``Two-loop five-point amplitudes of super Yang-Mills and supergravity in pure spinor superspace,''
JHEP {\bf 1510}, 124 (2015)
doi:10.1007/JHEP10(2015)124
[arXiv:1505.02746 [hep-th]].

\bibitem{Stieberger:2013wea} S.~Stieberger,
``Closed superstring amplitudes, single-valued multiple zeta values and the Deligne associator,''
J.\ Phys.\ A {\bf 47}, 155401 (2014)
doi:10.1088/1751-8113/47/15/155401
[arXiv:1310.3259 [hep-th]].

\bibitem{Mafra:2017ioj} C.~R.~Mafra and O.~Schlotterer,
``Double-Copy Structure of One-Loop Open-String Amplitudes,''
Phys.\ Rev.\ Lett.\  {\bf 121}, no. 1, 011601 (2018)
doi:10.1103/PhysRevLett.121.011601
[arXiv:1711.09104 [hep-th]].

\bibitem{Huang:2016tag} Y.~t.~Huang, O.~Schlotterer and C.~Wen,
``Universality in string interactions,''
JHEP {\bf 1609}, 155 (2016)
doi:10.1007/JHEP09(2016)155
[arXiv:1602.01674 [hep-th]].

\bibitem{Azevedo:2018dgo} 
T.~Azevedo, M.~Chiodaroli, H.~Johansson and O.~Schlotterer,
``Heterotic and bosonic string amplitudes via field theory,''
JHEP {\bf 1810}, 012 (2018)
doi:10.1007/JHEP10(2018)012
[arXiv:1803.05452 [hep-th]].

\bibitem{Cachazo:2013hca} F.~Cachazo, S.~He and E.~Y.~Yuan,
``Scattering of Massless Particles in Arbitrary Dimensions,''
Phys.\ Rev.\ Lett.\  {\bf 113}, no. 17, 171601 (2014)
doi:10.1103/PhysRevLett.113.171601
[arXiv:1307.2199 [hep-th]].

\bibitem{Cachazo:2013iea} F.~Cachazo, S.~He and E.~Y.~Yuan,
``Scattering of Massless Particles: Scalars, Gluons and Gravitons,''
JHEP {\bf 1407}, 033 (2014)
doi:10.1007/JHEP07(2014)033
[arXiv:1309.0885 [hep-th]].

\bibitem{Cachazo:2013gna} F.~Cachazo, S.~He and E.~Y.~Yuan,
``Scattering equations and Kawai-Lewellen-Tye orthogonality,''
Phys.\ Rev.\ D {\bf 90}, no. 6, 065001 (2014)
doi:10.1103/PhysRevD.90.065001
[arXiv:1306.6575 [hep-th]].

\bibitem{Cachazo:2014nsa} F.~Cachazo, S.~He and E.~Y.~Yuan,
``Einstein-Yang-Mills Scattering Amplitudes From Scattering Equations,''
JHEP {\bf 1501}, 121 (2015)
doi:10.1007/JHEP01(2015)121
[arXiv:1409.8256 [hep-th]].

\bibitem{Cachazo:2014xea} F.~Cachazo, S.~He and E.~Y.~Yuan,
``Scattering Equations and Matrices: From Einstein To Yang-Mills, DBI and NLSM,''
JHEP {\bf 1507}, 149 (2015)
doi:10.1007/JHEP07(2015)149
[arXiv:1412.3479 [hep-th]].

\bibitem{Bjerrum-Bohr:2016axv} N.~E.~J.~Bjerrum-Bohr, J.~L.~Bourjaily, P.~H.~Damgaard and B.~Feng,
``Manifesting Color-Kinematics Duality in the Scattering Equation Formalism,''
JHEP {\bf 1609}, 094 (2016)
doi:10.1007/JHEP09(2016)094
[arXiv:1608.00006 [hep-th]].

\bibitem{Mason:2013sva} L.~Mason and D.~Skinner,
``Ambitwistor strings and the scattering equations,''
JHEP {\bf 1407}, 048 (2014)
doi:10.1007/JHEP07(2014)048
[arXiv:1311.2564 [hep-th]].

\bibitem{Casali:2015vta} E.~Casali, Y.~Geyer, L.~Mason, R.~Monteiro and K.~A.~Roehrig,
``New Ambitwistor String Theories,''
JHEP {\bf 1511}, 038 (2015)
doi:10.1007/JHEP11(2015)038
[arXiv:1506.08771 [hep-th]].

\bibitem{deWit:1991nm} B.~de Wit and A.~Van Proeyen,
``Special geometry, cubic polynomials and homogeneous quaternionic spaces,''
Commun.\ Math.\ Phys.\  {\bf 149}, 307 (1992)
doi:10.1007/BF02097627
[hep-th/9112027].

\bibitem{Chiodaroli:2013upa} M.~Chiodaroli, Q.~Jin and R.~Roiban,
``Color/kinematics duality for general abelian orbifolds of N=4 super Yang-Mills theory,''
JHEP {\bf 1401}, 152 (2014)
doi:10.1007/JHEP01(2014)152
[arXiv:1311.3600 [hep-th]].

\bibitem{Bershadsky:1998cb} M.~Bershadsky and A.~Johansen,
``Large N limit of orbifold field theories,''
Nucl.\ Phys.\ B {\bf 536}, 141 (1998)
doi:10.1016/S0550-3213(98)00526-4
[hep-th/9803249].

\bibitem{Luciani:1977hp} J.~F.~Luciani,
``Coupling of O(2) Supergravity with Several Vector Multiplets,''
Nucl.\ Phys.\ B {\bf 132}, 325 (1978).
doi:10.1016/0550-3213(78)90123-2

\bibitem{Gunaydin:1984ak} M.~Gunaydin, G.~Sierra and P.~K.~Townsend,
``Gauging the d = 5 Maxwell-Einstein Supergravity Theories: More on Jordan Algebras,''
Nucl.\ Phys.\ B {\bf 253}, 573 (1985).
doi:10.1016/0550-3213(85)90547-4

\bibitem{Gunaydin:1983rk} M.~Gunaydin, G.~Sierra and P.~K.~Townsend,
``Exceptional Supergravity Theories and the MAGIC Square,''
Phys.\ Lett.\  {\bf 133B}, 72 (1983).
doi:10.1016/0370-2693(83)90108-9

\bibitem{Gunaydin:1983bi} M.~Gunaydin, G.~Sierra and P.~K.~Townsend,
``The Geometry of N=2 Maxwell-Einstein Supergravity and Jordan Algebras,''
Nucl.\ Phys.\ B {\bf 242}, 244 (1984).
doi:10.1016/0550-3213(84)90142-1

\bibitem{Elvang:2010jv} H.~Elvang, D.~Z.~Freedman and M.~Kiermaier,
``A simple approach to counterterms in N=8 supergravity,''
JHEP {\bf 1011}, 016 (2010)
doi:10.1007/JHEP11(2010)016
[arXiv:1003.5018 [hep-th]].

\bibitem{Bossard:2010bd} G.~Bossard, P.~S.~Howe and K.~S.~Stelle,
``On duality symmetries of supergravity invariants,''
JHEP {\bf 1101}, 020 (2011)
doi:10.1007/JHEP01(2011)020
[arXiv:1009.0743 [hep-th]].

\bibitem{Bossard:2011tq} G.~Bossard, P.~S.~Howe, K.~S.~Stelle and P.~Vanhove,
``The vanishing volume of D=4 superspace,''
Class.\ Quant.\ Grav.\  {\bf 28}, 215005 (2011)
doi:10.1088/0264-9381/28/21/215005
[arXiv:1105.6087 [hep-th]].

\bibitem{Beisert:2010jx} N.~Beisert, H.~Elvang, D.~Z.~Freedman, M.~Kiermaier, A.~Morales and S.~Stieberger,
``E7(7) constraints on counterterms in N=8 supergravity,''
Phys.\ Lett.\ B {\bf 694}, 265 (2011)
doi:10.1016/j.physletb.2010.09.069
[arXiv:1009.1643 [hep-th]].

\bibitem{Elvang:2010xn} H.~Elvang, D.~Z.~Freedman and M.~Kiermaier,
``SUSY Ward identities, Superamplitudes, and Counterterms,''
J.\ Phys.\ A {\bf 44}, 454009 (2011)
doi:10.1088/1751-8113/44/45/454009
[arXiv:1012.3401 [hep-th]].

\bibitem{Bossard:2012xs} G.~Bossard, P.~S.~Howe and K.~S.~Stelle,
``Anomalies and divergences in N=4 supergravity,''
Phys.\ Lett.\ B {\bf 719}, 424 (2013)
doi:10.1016/j.physletb.2013.01.021
[arXiv:1212.0841 [hep-th]].

\bibitem{Freedman:2011uc} D.~Z.~Freedman and E.~Tonni,
``The $D^{2k} R^4$ Invariants of N=8 Supergravity,''
JHEP {\bf 1104}, 006 (2011)
doi:10.1007/JHEP04(2011)006
[arXiv:1101.1672 [hep-th]].

\bibitem{Bossard:2013rza} G.~Bossard, P.~S.~Howe and K.~S.~Stelle,
``Invariants and divergences in half-maximal supergravity theories,''
JHEP {\bf 1307}, 117 (2013)
doi:10.1007/JHEP07(2013)117
[arXiv:1304.7753 [hep-th]].

\bibitem{Brodel:2009hu} J.~Broedel and L.~J.~Dixon,
``R**4 counterterm and E(7)(7) symmetry in maximal supergravity,''
JHEP {\bf 1005}, 003 (2010)
doi:10.1007/JHEP05(2010)003
[arXiv:0911.5704 [hep-th]].

\bibitem{Kallosh:2011dp} R.~Kallosh,
JHEP {\bf 1203}, 083 (2012)
doi:10.1007/JHEP03(2012)083
[arXiv:1103.4115 [hep-th]].

\bibitem{Kallosh:2011qt} R.~Kallosh,
``N=8 Counterterms and $E_{7(7)}$ Current Conservation,''
JHEP {\bf 1106}, 073 (2011)
doi:10.1007/JHEP06(2011)073
[arXiv:1104.5480 [hep-th]].

\bibitem{Kallosh:2012ei} R.~Kallosh,
``On Absence of 3-loop Divergence in N=4 Supergravity,''
Phys.\ Rev.\ D {\bf 85}, 081702 (2012)
doi:10.1103/PhysRevD.85.081702
[arXiv:1202.4690 [hep-th]].

\bibitem{Kallosh:2012yy} R.~Kallosh and T.~Ortin,
``New E77 invariants and amplitudes,''
JHEP {\bf 1209}, 137 (2012)
doi:10.1007/JHEP09(2012)137
[arXiv:1205.4437 [hep-th]].

\bibitem{Kalin:2018thp} G.~K\"{a}lin, G.~Mogull and A.~Ochirov,
``Two-loop $\mathcal{N}=2$ SQCD amplitudes with external matter from iterated cuts,''
arXiv:1811.09604 [hep-th].

\bibitem{ElvangHuang} Elvang, H. and Huang, Y.  ``Scattering Amplitudes in Gauge Theory and Gravity,'' Cambridge: Cambridge University Press (2015). doi:10.1017/CBO9781107706620


\end{thebibliography}
\end{document}